\documentclass{aa}
\usepackage{psfig}

\def\sbs{SBS 1129+576}

\def\h2{H{\small II}}

\newcounter{qub}
\begin{document}

\title{Spectroscopic and photometric studies of low-metallicity star-forming 
dwarf galaxies. I. SBS 1129+576}

\author{N. G.\ Guseva \inst{1}
\and P.\ Papaderos \inst{2}
\and Y. I.\ Izotov \inst{1}
\and R. F. Green \inst{3}
\and K. J.\ Fricke   \inst{2}
\and T. X.\ Thuan\inst{4}
\and K.G.\ Noeske\inst{2}}
\offprints{N. G. Guseva, guseva@mao.kiev.ua}
\institute{      Main Astronomical Observatory,
                 Ukrainian National Academy of Sciences,
                 Zabolotnoho 27, Kyiv 03680,  Ukraine
\and
                 Universit\"ats--Sternwarte, Geismarlandstra\ss e 11,
                 D--37083 G\"ottingen, Germany
\and
                 National Optical Astronomy Observatory, 
                 Tucson, AZ 85726, USA
\and
                 Astronomy Department, University of Virginia, 
                 Charlottesville, VA 22903, USA
}

\date{Received \hskip 2cm; Accepted}

\abstract{Spectroscopy and $V,I$ CCD photometry of the dwarf irregular 
galaxy SBS~1129+576 are presented for the first time.
The CCD images reveal a chain of compact H {\sc ii} regions within the 
elongated 
low-surface-brightness (LSB) component of the galaxy.
Star formation takes place mainly in two high-surface-brightness 
H {\sc ii} regions.
The mean $(V-I)$ colour of the LSB component in the 
 surface brightness interval $\mu_V$ between 23 and 
26  mag arcsec$^{-2}$ is relatively blue $\sim$0.56$\pm$0.03 mag,
as compared to the $(V-I)$ $\sim$0.9 -- 1.0 for  
 the majority of known dwarf irregular and blue compact dwarf
(BCD)  galaxies.
Spectroscopy shows that the 
galaxy is among the most metal-deficient galaxies with an oxygen 
abundance 12 + log (O/H) = 7.36 $\pm$ 0.10 in the brightest H {\sc ii} region and
7.48 $\pm$ 0.12 in the second  brightest H {\sc ii} region, 
or 1/36 and 1/28 of the solar value\thanks{12+log(O/H)$_{\odot}$ = 8.92
(Anders \& Grevesse \cite{Anders89}).}, respectively. 
H$\beta$ and H$\alpha$ emission lines and H$\delta$ and H$\gamma$ absorption
lines are detected in a large part of the LSB component.
We use two extinction-insensitive methods
based on the equivalent widths of (1) emission and (2) absorption Balmer lines
to put constraints on the age of the stellar populations in the galaxy.
In addition, we use two extinction-dependent methods based on (3) the spectral 
energy distribution (SED) and (4) the $(V-I)$ colour. 
 Several scenarios of star formation were explored using all 4 methods.
The observed properties of the LSB component can be 
reproduced by a stellar population forming continuously 
since 10 Gyr ago, provided that the star formation rate has 
increased during the last 100 Myr by  a factor of 6 
to 50 and no extinction is present.
However, the observational properties of the LSB component in SBS 1129+576 can be 
reproduced equally well by continuous 
star formation which started not earlier than 100 Myr ago and stopped at 
5 Myr, if some extinction is assumed.
Hence, the ground-based spectroscopic and photometric observations are
not sufficient for distinguishing between a young and an old age for 
SBS 1129+576.
}


\maketitle

\markboth {N. G. Guseva et al.}{Spectroscopic and photometric studies of 
low-metallicity star-forming dwarf galaxies. I. SBS 1129+576}

\section{Introduction}

  \begin{figure*}
\vspace{1.cm}
    \hspace*{-0.0cm}\psfig{figure=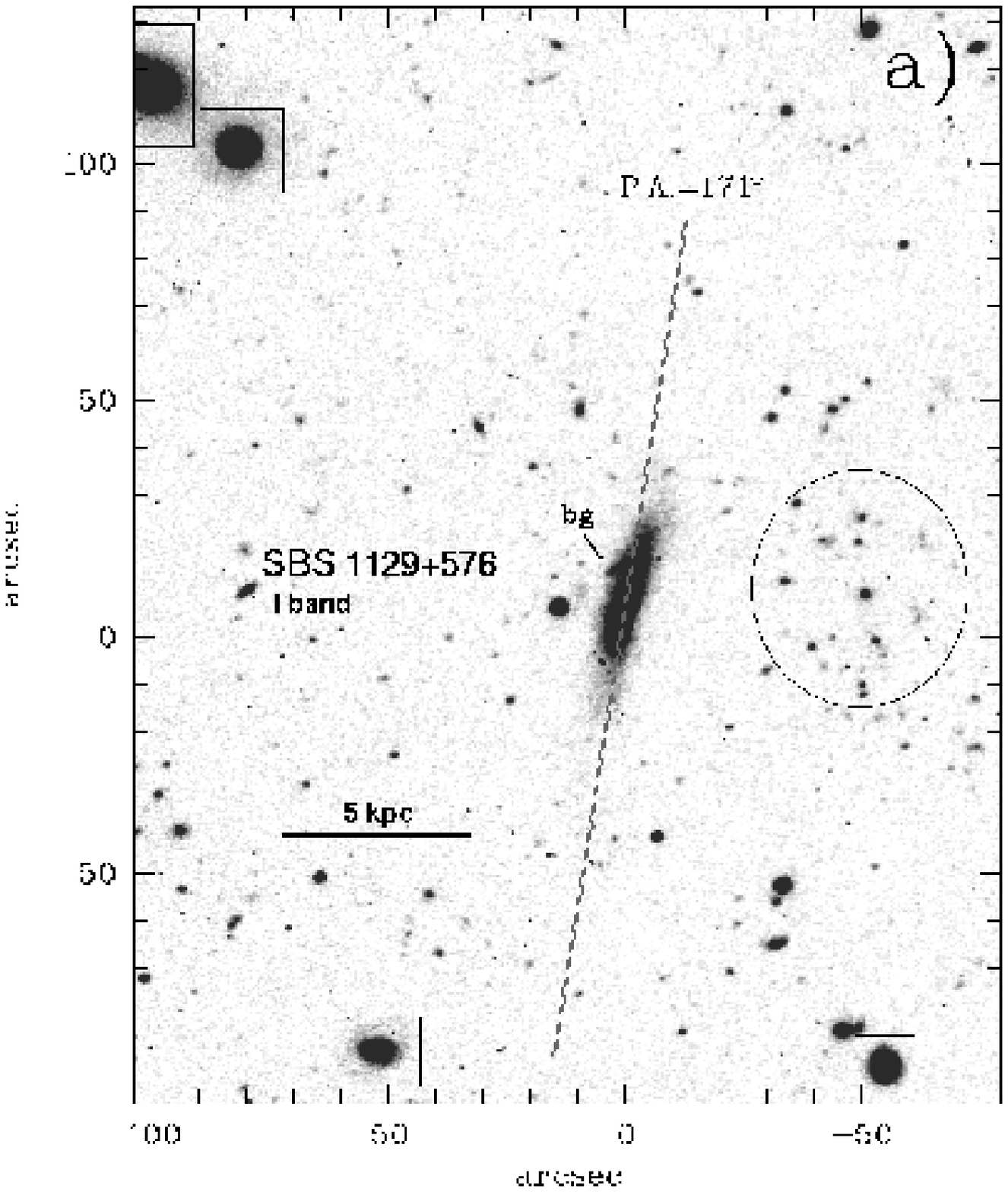,angle=0,height=9.2cm,clip=}
    \hspace*{0.2cm}\psfig{figure=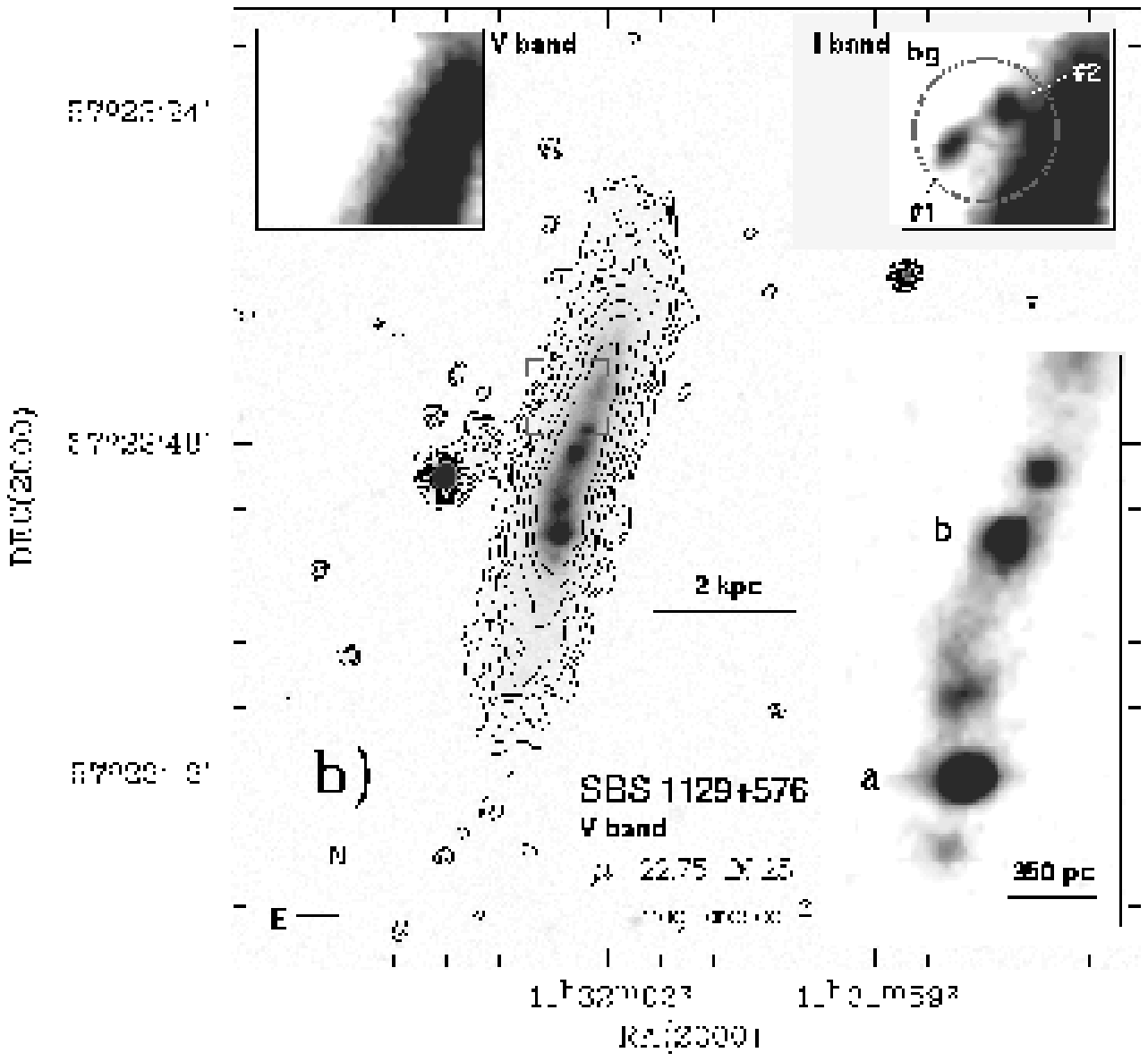,angle=0,height=9.2cm,clip=}
   \caption[]{({\bf a}) $I$ image of the field around 
SBS 1129+576. The straight line shows
the position of the slit during spectroscopic observations. A pair of red
background galaxies intersecting the LSB component 
 $\sim$15\arcsec\ northwards of the brightest region {\it a} 
(see panel {\bf b}) is labeled {\it bg}. 
The ellipse delineates the central part 
of a probable uncatalogued background galaxy cluster. 
Rectangles mark the position of some other 
uncatalogued galaxies in the field. These objects span a $(V-I)$ colour 
range between 1.1 and 1.3 mag. 
({\bf b}) $V$ image of SBS 1129+576. The overlayed contours correspond to 
surface brightness between 22.75
and 26.25 mag 
arcsec$^{-2}$ in steps of 0.5 mag. The inset to the lower-right, computed 
from the $V$ image after subtraction of a two-dimensional model for the LSB 
component, illustrates the spatial distribution of compact star-forming 
regions. The insets to the upper left and 
upper right show, respectively, close-up views in the $V$ and $I$ of the 
region depicted by rectangular brackets.  This 
region shows a smooth morphology in the $V$, while the $I$ image reveals 
two red ($(V-I)$$\sim$1.6) background galaxies labeled \#1 and \#2.
}    
\label{f1}
\end{figure*}

SBS 1129+576 ($\alpha$(J2000.0) = 11$^{\rm h}$32$^{\rm m}$02\fs5,
$\delta$(J2000.0) = +57$^\circ$22\arcmin45\farcs7, Bicay et al. 2000) 
was discovered in the course of the Second Byurakan Survey (SBS) 
(Markarian \& Stepanian 1983; Lipovetsky et al. 1988) as a galaxy with strong 
emission lines, weak continuum and ultraviolet excess seen in a chain of 
H {\sc ii} regions embedded within an extended blue 
low-surface-brightness (LSB) component.
Up to now SBS 1129+576 has not been studied in detail.
The low metallicity and relatively blue colour of its LSB component
(this paper) make it one of the rare young dwarf galaxy candidates
(Izotov $\&$ Thuan 1999).
In the present paper  the physical conditions and 
chemical abundances of the ionized gas of \sbs\
are studied for the first time.
In addition, spectroscopic and $V,I$ photometric
data are used to study the properties of the unresolved stellar population 
in its bright H {\sc ii} regions and  LSB component. 
Recently Thuan et al. (1999) derived for \sbs\ a redshift $z$ = 0.00522 from
single-dish H {\sc i} 21 cm observations.
After correction of the radial velocity for the Virgocentric 
infall motion, they derive 
a distance of $D$ = 26.3 Mpc, which we adopt. At this distance 1 arcsec
corresponds to a linear scale of 127 pc.

The structure of the paper is as follows. In Sect. \ref{obs} we describe the 
observations and data reduction. The photometric properties of SBS 1129+576 
are described in Sect. 3. In Sect. \ref{chem} we derive the chemical abundances in 
the two brightest H {\sc ii} regions. The properties of the stellar LSB 
population and its possible age range are discussed in Sect. \ref{age}.
Finally, Sect. \ref{conc} summarises the main conclusions of this work.

\section{Observations and data reduction \label{obs}}

\subsection{Photometric observations and data reduction}

Direct images of SBS 1129+576 in $V$ and $I$ (Fig. \ref{f1}) were acquired 
with the Kitt Peak 2.1m telescope\footnote{Kitt Peak National 
Observatory (KPNO) 
is operated by the Association of Universities for Research in 
Astronomy (AURA), 
    Inc., under cooperative agreement with the National Science 
Foundation (NSF).} 
on April 19, 1999, during a photometric night.
The telescope was equipped with a Tektronix 1024$\times$1024 CCD 
detector operating 
at a gain of 3\,e$^-$\,ADU$^{-1}$, giving an instrumental 
scale of 0\farcs305 pixel$^{-1}$ and field of view of 5\arcmin. 
The total exposures of 20 and 30 min in $V$ and $I$, 
respectively, were split into four subexposures each being slightly offset 
with respect to each other for removal of cosmic particle hits and 
bad pixels. 
The point spread function in $V$ and $I$ were respectively
1\farcs06 and 1\farcs17 FWHM.
Bias and flat--field images were obtained at the beginning and end
of night. 
Calibration was achieved by observing 4 standard 
fields from Landolt (\cite{Landolt92}) at 3--4 different airmasses
during the night.
Our calibration uncertainties are estimated to be 0.01--0.02 mag 
in each of the $V$ and $I$ bands.
The data reduction, including bias subtraction, 
removal of cosmic particle hits,
flat--field 
correction and absolute flux calibration was made using  
IRAF\footnote{IRAF is the Image 
Reduction and Analysis Facility distributed by the  
National Optical Astronomy Observatory, which is operated by the 
AURA under cooperative agreement with the NSF.}.

\subsection{Spectroscopic observations and data reduction}

 The spectroscopic observations were carried out on 19 June, 1999, with the Kitt Peak 4m 
Mayall telescope and Ritchey-Chretien spectrograph with the 
T2KB 2048~$\times$~2048 CCD detector, with the slit at P.A. = 171$^{\circ}$, 
centered on the brightest star-forming region and extending along the 
elongated body of the galaxy (roughly aligned with the major axis; 
see Fig.~\ref{f1}a).
 A 2\arcsec~$\times$~300\arcsec ~slit with the KPC-10A grating in 
first order and a GG 375 order separation filter was used.
The spatial scale along the slit
was 0\farcs69 pixel$^{-1}$ and the spectral resolution $\sim$7~\AA\ (FWHM).
The spectra were obtained at an airmass 1.33 and in a total 
exposure of 60 minutes, which was broken up into 3 subexposures. 
No correction for atmospheric refraction was made because the slit was 
oriented with a P.A. close to
the parallactic angle. 
Two Kitt Peak 
spectrophotometric standard stars were observed for flux calibration.
For wavelength calibration, spectra of a He-Ne-Ar comparison lamp were taken
after each exposure.

The data reduction was made with the IRAF software package. This includes 
bias--subtraction, flat--field correction, cosmic-ray removal, wavelength 
calibration, night sky subtraction, correction for atmospheric 
extinction and absolute flux calibration of the two--dimensional spectra.

One-dimensional spectra for abundance determination in
the two brightest H {\sc ii} regions {\it a} 
and {\it b} (Figs.~\ref{f1}b and \ref{fig:SBS1129map2})
were extracted from the two-dimensional spectrum within large apertures of 
2\arcsec\ $\times$ 5\arcsec.  
Some additional spectra of regions {\it a} and {\it b}  within smaller apertures 
2\arcsec~$\times$ 1\farcs1,  2\arcsec~$\times$ 1\farcs4 and 
2\arcsec~$\times$ 2\farcs1 
were also extracted. 

In addition we extracted spectra showing hydrogen Balmer absorption lines
for five regions along  the major axis of the galaxy
to study the stellar population of the LSB component.
The locations of selected regions, denoted 1 to 5, relative to region {\it a}
are given
in Tables \ref{t:emhahb} and \ref{t:abshdhg}. 
The spatial extent of these regions along the slit were 6\farcs2, 5\farcs5, 
4\farcs8, 4\farcs8 and 4\farcs8, respectively.

\section{Photometric analysis}
\subsection{Morphology, environment and colour distribution \label{morph}}

\begin{table*}
\caption{\label{decomp_res}Structural properties of the starburst and LSB components of SBS 1129+576$^{\rm a}$.}
\label{photom}
\begin{tabular}{lcccccccccc}
\hline\hline
Band & $\mu_{\rm E,0}$ & $\alpha $ & $P_{25}$  & $m_{P_{25}}$ & $E_{25}$ & $m_{\rm
  E_{25}}$ & $m_{\rm LSB}$  & $m_{\rm SBP}$ & $m_{\rm tot}$ & $r_{\rm eff}$,$r_{80}$  \\
          & mag arcsec$^{-2}$ & pc       & kpc       &  mag             &  kpc
          &  mag             & mag      &   mag & mag & kpc \\
    (1) &   (2)            &   (3)     &  (4)      &   (5)            &
 (6)    &  (7)             &  (8)     &  (9)  & (10) & (11) \\
\hline
 $V$ & 21.01$\pm$0.02  & 436$\pm$4  & 0.73  & 18.27  & 1.59  & 16.54  & 16.43 &
 16.23$\pm$0.01 & 16.22 & 0.65,1.17\\
 $I$ & 20.43$\pm$0.03  & 433$\pm$4  & 0.78  & 18.25  & 1.82  & 15.92  & 15.86 &
 15.73$\pm$0.02  & 15.72 & 0.69,1.20\\
\hline
\end{tabular}

$^{\rm a}$The tabulated values have not been corrected for interstellar extinction or 
inclination.
\end{table*}

\begin{figure}[hbtp]
\vspace{1.cm}
    \hspace{0cm}\psfig{figure=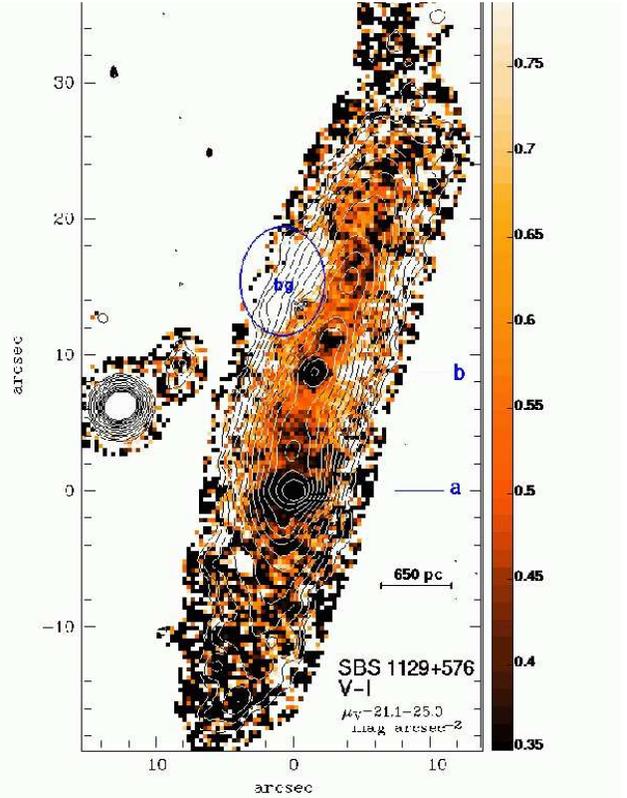,angle=0,width=8.4cm,clip=}
    \caption{ $(V-I)$ map of SBS 1129+576 displayed in the colour 
range between +0.35 to +0.8 mag. The overlayed contours are from 21.1 to 25.0 
$V$ mag arcsec$^{-2}$ in steps of 0.3 mag. 
The region, depicted by the ellipse, shows a strikingly redder $(V-I)$ colour 
($\sim$ 1.6 mag) than the average colour $\sim$ 0.56 mag of the LSB component. 
This local maximum in the $(V-I)$ colour is to be attributed to background 
galaxies labeled {\it bg} in Fig. \ref{f1}a and in the upper-right inset of Fig. 
\ref{f1}b.
      }
    \label{fig:SBS1129map2}
\end{figure}

   \begin{figure*}[hbtp]
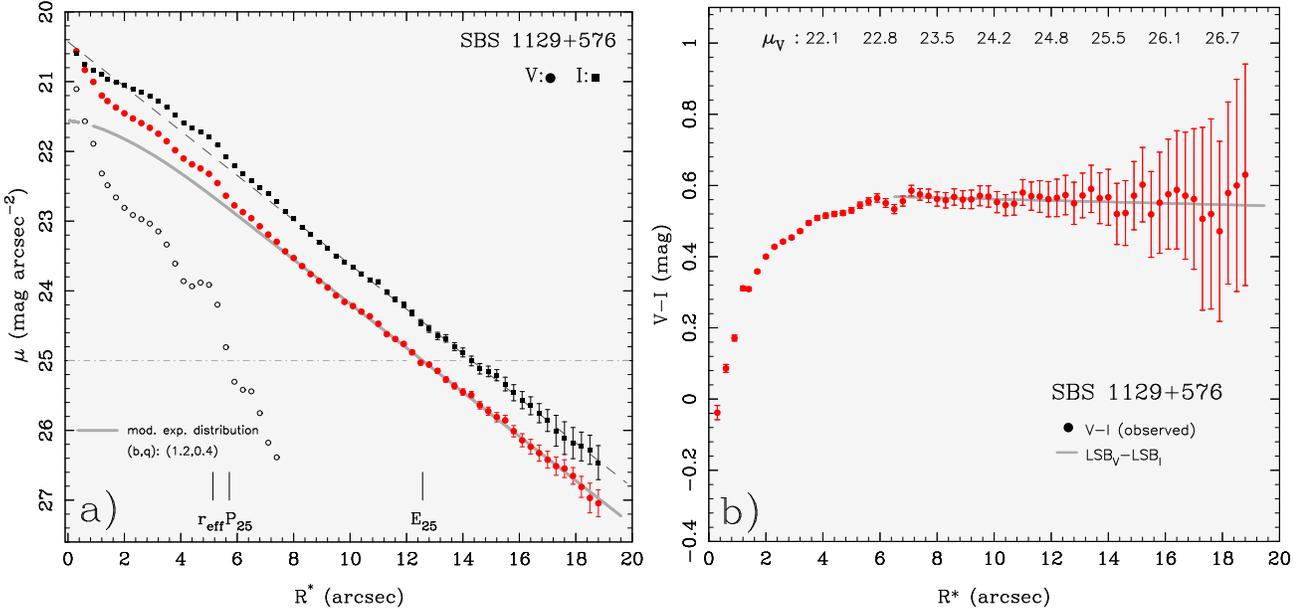

   \hspace*{-0.0cm}\psfig{figure=3327.f3a.ps,angle=270,width=8.5cm}
   \hspace*{-0.0cm}\psfig{figure=3327.f3b.ps,angle=270,width=8.5cm}
\caption[]{{\bf (a)} Surface brightness profiles (SBP) of SBS 1129+576
in the $V$ and $I$ derived using the method iii.
The modeled surface brightness distribution of the 
 LSB component in the $V$ assuming a modified exponential distribution 
with ($b$,$q$)=(1.2,0.4) 
is shown by the thick-grey curve. 
Open circles show the surface brightness distribution 
of the light 
in excess to the modeled LSB profile
and is to be attributed to the compact star-forming
regions along the major axis of the galaxy (Fig. \ref{f1}b). 
The effective radius $r_{\rm eff}$ of the $V$ SBP
and the isophotal radii $P_{25}$ and $E_{25}$ of, respectively, 
the starburst and the LSB component at 25 $V$ mag arcsec$^{-2}$ 
are indicated. 
The dashed line shows a linear fit to the $I$ band profile
for $R^*\geq$8\arcsec, extrapolated to $R^*$ = 0.
{\bf (b)} Circles show the ~$(V-I)$ colour profile 
computed by subtraction of the SBPs displayed in the left panel. 
Labels attached to the upper axis indicate the $V$ surface brightness 
corresponding to the respective photometric radius $R^*$.
The $(V-I)$ colour distribution (thick-grey line) is
obtained by subtraction of the 
fits in $V$ and $I$ shown in {\bf (a)}.
It closely matches the observed colour distribution and 
implies a practically constant  $(V-I)$ colour of $\approx$ 0.56 mag 
over a range of 4 magnitudes in 
surface brightness.
}
\label{f2}
\end{figure*}

SBS 1129+576 appears as an oblate dwarf ($M_V$ $\sim$ --15.9 mag) galaxy with 
a projected axial ratio of 4 at 25 $V$ mag arcsec$^{-2}$ (Fig. \ref{f1}, 
\ref{fig:SBS1129map2}). 
Star formation activity takes place in the central parts of the galaxy
within an elongated, moderately blue region 
with a size of $\sim$ 2.5 kpc.
The $V$ magnitudes of regions {\it a} and {\it b} are 
respectively $\approx$ 19.0 mag and 
20.0 mag, corresponding to absolute $V$ magnitudes --13.1 mag and --12.1 mag.
The $(V-I)$ colours of $\sim$ 0.1 mag and $\sim$ 0.3 mag, respectively,
of regions {\it a} and {\it b} are relatively blue. 
The $(V-I)$ colour of the LSB component shows a 
smooth change from $\sim$ 0.44 mag in the immediate vicinity of region 
{\it a} to an average colour of $\sim$ 0.56 $\pm$ 0.03 mag 
in the outer part
of the galaxy (see Fig. \ref{f2}b).
The $(V-I)$ colour map (Fig. \ref{fig:SBS1129map2}) reveals for surface 
brightness levels fainter than 24 $V$ mag arcsec$^{-2}$, a featureless and 
relatively constant colour over the whole LSB component 
except for a strikingly red ($(V-I)$ $\sim$ 1.6 mag) region located 
$\sim$ 15\arcsec\ north of region {\it a} (region $bg$ 
in Figs.~\ref{f1} and~\ref{fig:SBS1129map2}).
The local colour excess observed in region $bg$ is due 
to two background galaxies seen in  the
$I$ image only. A close-up view of this region in the $V$ and $I$
is shown in the upper insets of 
Fig. \ref{f1}b. 
A potential slight overestimate of the $(V-I)$ colour of the LSB component in
SBS 1129+576 as a result of the superposition of background sources 
at different locations is likely given the numerous red 
($(V-I)$ $>$ 1.2 mag) faint ($m_I$ $\ga$ 20.5 mag) sources in the 
field of the galaxy. 
Figure~\ref{f1}a shows that  SBS 1129+576 is located in front of a 
probable background 
cluster of galaxies centered $\sim$ 50\arcsec\ west 
of region {\it a}.
This cluster with the central part delineated by the ellipse 
as well as several red galaxies indicated by rectangles in 
Fig. \ref{f1}a are not catalogued in the NASA/IPAC 
Extragalactic Database (NED).

\subsection{Surface photometry}
Surface brightness profiles (SBPs) of \sbs\ have been computed following the 
methods i through iii described in Papaderos et al. (\cite{Papa96a}). 
Briefly, the photometric radius
$R^*=(A(\mu)/\pi)^{0.5}$ corresponding to the surface brightness level
$\mu$ is computed from the area $A(\mu)$ of a galaxy in arcsec$^2$, 
as derived through ellipse fitting or computation of a line-integral along 
an isophote (methods i and ii) or summation of all pixels inside a polygonal aperture
with a surface brightness brighter than $\mu$ (method iii). 
Essentially, these techniques trace the growth of the isophotal 
size of a galaxy with decreasing intensity $I$. 
They require no choice of a ``geometrical center'' of a galaxy
and insure that the photometric radius $R^*$ is a monotonic function of $\mu$.
Evidently, in order to derive SBPs as described above, one has to keep track of 
the morphology of a BCD throughout its intensity span, i.e. in general to be 
able to interpolate an isophote down to the faintest measured level $\mu$ of a
SBP. This allows to visually check for and screen-out foreground and 
background sources in the periphery of the galaxy, thus to make sure 
that source confusion does not affect the SBP slope at faint intensity levels.
This task is more difficult to achieve when computing SBPs 
based on photon statistics inside circular or elliptical 
annuli, extending out to a user-defined maximal radius $r_{\rm max}$. 
Especially for BCDs, SBPs derived with the latter methods may considerably
vary, depending on subjective assumptions on the $r_{\rm max}$ or the
``center'' of a galaxy.

As a check for consistency, we also computed SBPs using method iv in 
Papaderos et al. (\cite{papaderos02}). This technique is based on the calculation
of photon statistics for a series of masks of arbitrary (generally irregular) 
shape, mapping equidistant logarithmic intensity intervals between $I_{\rm min}$ 
and $I_{\rm max}$. As in methods i through iii, method iv 
does not require the choice of a ``geometrical center'' and accounts 
adequately for the large morphological variation of a BCD,
with a typically smooth LSB part and an irregular star-forming component.

We derived $V$ and $I$ SBPs from the \emph{total} light of the galaxy,
except for region {\it bg} (shown by the ellipse in
Fig. \ref{fig:SBS1129map2}) which has been replaced by a two-dimensional fit to the adjacent LSB emission. 
Both SBPs in Fig. \ref{f2}a are derived using the method iii. They 
show an exponential intensity decrease for radii
$R^*\ga 8$\arcsec\ with a scale length $\alpha\approx$3\farcs4.
That the exponential model yields a reasonable approximation to the LSB emission
is also indicated from fitting a S\'ersic (\cite{sersic68}) profile
\begin{equation}
I(R^*) = I_0\,\exp\left(-\frac{R^*}{\beta}\right)^{\eta}
\label{eq:sersic} 
\end{equation}
(see also Caon et al. \cite{Caon93}, Cellone et al. \cite{cellone94},
Papaderos et al. \cite{Papa96a}) to the radius interval $R^*\geq$8\arcsec.
The exponent $\eta$ obtained this way, respectively 1.2 and 1.1 in $V$ and 
$I$, is marginally larger than the value $\eta$ = 1, 
corresponding to the exponential law.

However, a closer inspection of the $I$ band SBP (Fig. \ref{f2}a)
shows that an inwards extrapolation of the exponential law fitted to 
the LSB component (see dashed line, overlayed with the $I$ profile)
predicts for small radii ($R^*$$\la$2\arcsec) a slightly higher intensity 
than the one observed. Such a pure exponential LSB model would imply 
that the star-forming component in \sbs\ provides no more than 5\% of the 
total $I$ emission and that its contribution decreases rapidly at small radii.
This suggests that the stellar LSB emision of \sbs\ is best approximated by an
exponential profile with a flattening in its inner part. Note that such SBPs
have been frequently observed in dwarf ellipticals 
(e.g., Binggeli \& Cameron \cite{binggeli91}), 
dwarf irregulars (R\"onnback \& Bergvall 1994; Patterson \& Thuan 
\cite{patterson96}; Makarova \cite{m99}; van Zee \cite{v2000}) and a few 
blue compact dwarf (BCD) galaxies (e.g., 
Papaderos et al. \cite{Papa96a}; 
Vennik et al. \cite{vennik96}, \cite{vennik00}; Telles \& Terlevich \cite{telles97}; 
Guseva et al. \cite{Guseva2001};
Fricke et al. \cite{Fricke00}).
SBPs of this kind, classified ``type~V'' in Binggeli \& Cameron (\cite{binggeli91}),
can be approximated by, e.g., the modifed exponential distribution proposed in 
Papaderos et al. (\cite{Papa96a}):
\begin{equation}
\label{med}
I(R^{\star})=I_{\rm E,0}\exp
\left(-\frac{R^{\star}}{\alpha}\right)\{1-q\exp (-P_3(R^{\star}))\}
\end{equation}
with
\begin{equation}
P_3(R^{\star})=\left(\frac{R^{\star}}{b\alpha}\right)^3+
\left(\frac{R^{\star}}{\alpha} \frac{1-q}{q}\right).
\end{equation}
The empirical fitting function (Eq. \ref{med}) flattens with respect 
to the exponential law inside of a cutoff radius $b\alpha$, and attains at
$R^*$=0\arcsec\ an intensity given by the relative depression parameter
$q=\Delta I/I_{\rm E,0}<1$. An advantage of Eq. \ref{med} is that its
exponential part, depending on $I_{E,0}$ and 
$\alpha$ only, can be readily constrained from linear fits to  
the outer exponential part of a ``type~V'' SBP.

In order to disentangle the intensity distribution of the LSB component 
and better constrain the depression parameters $b$ and $q$ in Eq. \ref{med}
we follow the approach adopted in Guseva et al. (\cite{Guseva2001}).
We first subtracted compact (diameter $\la$4\arcsec) high-surface-brightness regions 
in the inner part of the galaxy and then rederived SBPs from
the residual underlying LSB emission.
Fitting Eq. \ref{med} to the resulting profiles
yields $V$ and $I$ depression parameters ($b$,$q$) = (1.2,0.4) 
and an exponential scale length $\alpha$ $\sim$ 430 pc. 
In Fig. \ref{f2}a we show for the $V$ SBP the modeled 
surface brightness distribution of the LSB component according to Eq. \ref{med}
and the emission in excess of the model with the thick-grey 
curve and open circles, respectively. 
The dashed line shows a linear fit to the $I$ band profile
for $R^*\geq$8\arcsec, extrapolated to $R^*$ = 0.
The excess emission is due to the chain of 
compact star-forming regions along the major axis of the galaxy.

\begin{figure*}[hbtp]
   \hspace*{2.5cm}\psfig{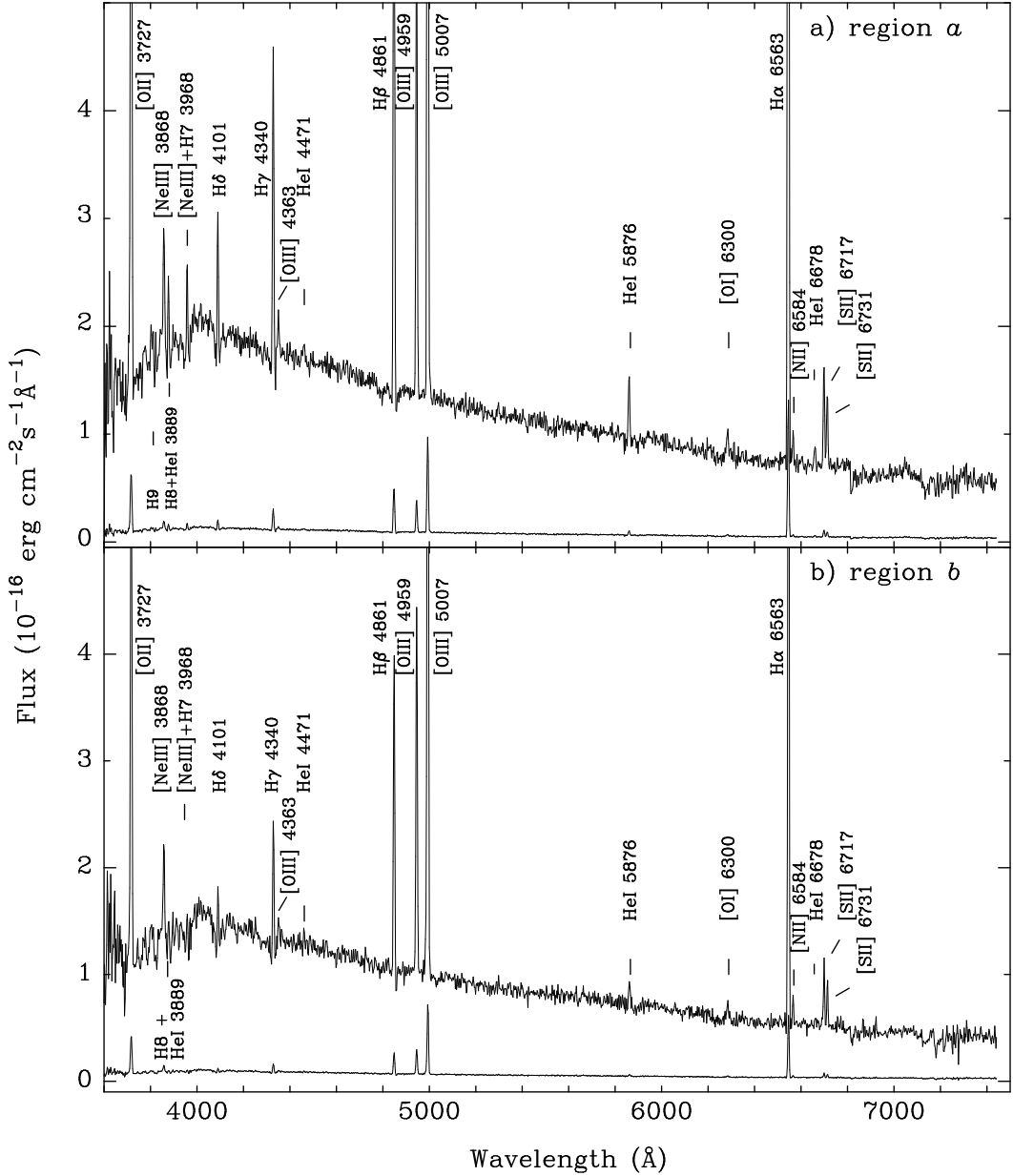}
    \caption{The KPNO 4m spectra of regions {\it a} and {\it b} with 
    the emission lines identified.
     The lower spectra in ({\bf a}) and ({\bf b}) are the observed 
     spectra downscaled by a factor of 15.
      }
    \label{fig:brightsp}
\end{figure*}

Table \ref{photom} summarizes the derived photometric quantities.
Cols.\ 2 and 3 give, respectively, the central surface brightness 
$\mu_{\rm E,0}$ and scale length $\alpha$ of the LSB component as obtained 
from linear fits to the SBPs for $R^*\geq 8$\arcsec\ and weighted by 
the photometric uncertainty of each point. 
These quantities correspond to the values one would obtain 
from extrapolation of the exponential slope observed in the 
outlying regions of the galaxy to $R^* = 0$.  
Cols. 4 through 8 list quantities obtained from profile decomposition 
where the intensity distribution of the LSB component was modeled by 
the modified exponential distribution (Eq. \ref{med}).
Cols.\ 4 and 6 give the radial extent $P_{25}$ and $E_{25}$
of the starburst and LSB components, respectively, both determined 
at a surface brightness  25\ mag arcsec$^{-2}$.
The apparent magnitudes of both components within $P_{25}$ and $E_{25}$
are listed in cols.\ 5 and 7, respectively. 
Col.\ 8  gives the apparent magnitude of 
the LSB component in each band within a photometric radius 
of 18\arcsec, as derived from integration of the modeled 
distribution (Eq. \ref{med}). The total magnitudes of the galaxy,
as inferred from SBP integration out to the same radius and by
integrating the flux within a polygonal aperture are listed in
cols. 9 and 10, respectively. Col.\ 11 gives the effective 
radius $r_{\rm eff}$ and the radius $r_{80}$ which encircles 
80\% of the galaxy's total flux.

From Table \ref{photom} it is evident that the integrated emission of the 
starburst component including the two 
brightest regions {\it a} and {\it b} contributes only  $\sim$ 17\% 
of the $V$ light of \sbs\ within its 25 $V$ mag arcsec$^{-2}$ isophote. 
This is a factor of 3 lower than the average of 50\% 
derived for BCDs in the $B$ band by Papaderos et al. (\cite{Papa96b}) and 
Salzer \& Norton (\cite{SN99}). 

The $(V-I)$ colour profile of SBS 1129+576, derived from subtraction 
of the corresponding SBPs, is shown in Fig. \ref{f2}b. 
Its behaviour is similar to that
in BCDs (Papaderos et al. \cite{Papa96a}) 
and compact irregular dwarf
galaxies (Patterson $\&$ Thuan \cite{patterson96}; van Zee \cite{v2000}) with a blue 
colour in the inner part of the galaxy and a redder, relatively constant 
colour at larger radii.
The colour in \sbs\ increases gradually 
from $(V-I)$ $\la$ 0.2 mag for radii $R^*$ $\la$ 2\arcsec\ to 
$\sim$ 0.5 mag at $R^*$ $\la$ 5\arcsec\ and remains practically
constant at $\sim$ 0.56 $\pm$ 0.03 mag in the outer part of the galaxy.  
The observed colour is in accord with the one resulting from subtraction of
the apparent magnitudes of the modeled distributions for the LSB component 
(Table \ref{photom}, col. 8) which has an average value of 0.57 mag.

\section{Chemical abundances \label{chem}}

In this Section we analyze the element abundances in 
SBS 1129+576 based on  spectroscopic observations
of the two brightest H {\sc ii} regions {\it a} and {\it b}. 
The spectra of these star-forming regions
(Fig. \ref{fig:brightsp}) 
are characterised by strong nebular emission lines superposed on 
stellar Balmer absorption lines. 
The latter are also seen along the slit in the LSB component.

\begin{table*}[tbh]
\caption{Observed ($F$($\lambda$)) and corrected  
($I$($\lambda$)) fluxes and equivalent widths ($EW$) of emission lines
 in regions {\it a} and {\it b}.}
\label{t:Intens}
\begin{tabular}{lccrcccr} \hline \hline
  &\multicolumn{3}{c}{region {\it a}}&&\multicolumn{3}{c}{region {\it b}} \\ \cline{2-4} \cline{6-8}
$\lambda_{0}$(\AA) Ion                  &$F$($\lambda$)/$F$(H$\beta$)
&$I$($\lambda$)/$I$(H$\beta$)&$EW$(\AA)&&$F$($\lambda$)/$F$(H$\beta$)&$I$($\lambda$)/$I$(H$\beta$) 
&$EW$(\AA)  \\ \hline
3727\ [O {\sc ii}]             & 1.408 $\pm$0.035 &  1.346 $\pm$0.037 & 51.0 $\pm$0.8 && 1.875 $\pm$0.061 & 1.758 $\pm$0.065 &  38.8 $\pm$0.7  \\
3835\ H9                       & 0.030 $\pm$0.013 &  0.084 $\pm$0.048 &  1.0 $\pm$0.4 &&       ...        &       ...        &        ...        \\
3868\ [Ne {\sc iii}]           & 0.169 $\pm$0.015 &  0.161 $\pm$0.015 &  4.5 $\pm$0.4 && 0.313 $\pm$0.026 & 0.293 $\pm$0.026 &   5.3 $\pm$0.4  \\
3889\ H8\ +\ He {\sc i}        & 0.128 $\pm$0.014 &  0.179 $\pm$0.026 &  4.2 $\pm$0.5 && 0.147 $\pm$0.025 & 0.240 $\pm$0.051 &   3.1 $\pm$0.5  \\
3968\ [Ne {\sc iii}]\ +H7      & 0.122 $\pm$0.012 &  0.177 $\pm$0.024 &  3.7 $\pm$0.4 && 0.158 $\pm$0.027 & 0.249 $\pm$0.052 &   3.3 $\pm$0.6  \\
4101\ H$\delta$                & 0.183 $\pm$0.013 &  0.235 $\pm$0.023 &  5.5 $\pm$0.4 && 0.180 $\pm$0.020 & 0.252 $\pm$0.044 &   3.5 $\pm$0.4  \\
4340\ H$\gamma$                & 0.457 $\pm$0.018 &  0.490 $\pm$0.024 & 15.6 $\pm$0.5 && 0.434 $\pm$0.026 & 0.483 $\pm$0.038 &   9.1 $\pm$0.5  \\
4363\ [O {\sc iii}]            & 0.061 $\pm$0.014 &  0.058 $\pm$0.014 &  1.7 $\pm$0.4 && 0.117 $\pm$0.023 & 0.110 $\pm$0.023 &   2.2 $\pm$0.4  \\
4471\ He {\sc i}               & 0.033 $\pm$0.014 &  0.032 $\pm$0.014 &  1.0 $\pm$0.4 && 0.042 $\pm$0.019 & 0.038 $\pm$0.019 &   0.8 $\pm$0.3  \\
4861\ H$\beta$                 & 1.000 $\pm$0.027 &  1.000 $\pm$0.030 & 41.0 $\pm$0.8 && 1.000 $\pm$0.040 & 1.000 $\pm$0.045 &  25.4 $\pm$0.8  \\
4959\ [O {\sc iii}]            & 0.666 $\pm$0.022 &  0.637 $\pm$0.022 & 24.4 $\pm$0.6 && 1.173 $\pm$0.044 & 1.099 $\pm$0.044 &  26.6 $\pm$0.7  \\
5007\ [O {\sc iii}]            & 1.975 $\pm$0.045 &  1.888 $\pm$0.045 & 73.8 $\pm$0.9 && 3.252 $\pm$0.099 & 3.049 $\pm$0.099 &  76.9 $\pm$1.0  \\
5876\ He {\sc i}               & 0.086 $\pm$0.010 &  0.082 $\pm$0.010 &  4.5 $\pm$0.5 && 0.093 $\pm$0.017 & 0.087 $\pm$0.017 &   3.0 $\pm$0.6  \\
6300\ [O {\sc i}]              & 0.048 $\pm$0.010 &  0.046 $\pm$0.010 &  3.0 $\pm$0.6 && 0.075 $\pm$0.017 & 0.069 $\pm$0.017 &   3.1 $\pm$0.7  \\
6563\ H$\alpha$                & 2.692 $\pm$0.058 &  2.599 $\pm$0.064 &186.3 $\pm$1.8 && 2.662 $\pm$0.079 & 2.526 $\pm$0.088 & 137.3 $\pm$1.6  \\
6584\ [N {\sc ii}]             & 0.041 $\pm$0.008 &  0.039 $\pm$0.008 &  2.8 $\pm$0.5 && 0.071 $\pm$0.013 & 0.066 $\pm$0.013 &   3.1 $\pm$0.6  \\
6678\ He {\sc i}               & 0.030 $\pm$0.008 &  0.029 $\pm$0.008 &  2.1 $\pm$0.6 && 0.026 $\pm$0.015 & 0.024 $\pm$0.015 &   1.1 $\pm$0.7  \\
6717\ [S {\sc ii}]             & 0.142 $\pm$0.010 &  0.136 $\pm$0.010 & 10.1 $\pm$0.7 && 0.168 $\pm$0.018 & 0.155 $\pm$0.018 &   7.5 $\pm$0.8  \\
6731\ [S {\sc ii}]             & 0.103 $\pm$0.010 &  0.098 $\pm$0.010 &  7.3 $\pm$0.7 && 0.114 $\pm$0.015 & 0.106 $\pm$0.015 &   5.1 $\pm$0.6  \\
                     & & & & & & & \\
$C$(H$\beta$)\ dex             &\multicolumn {3}{c}{0.000$\pm$0.028} &&\multicolumn {3}{c}{0.000$\pm$0.041} \\
$F$(H$\beta$)$^{\rm a}$          &\multicolumn {3}{c}{0.50$\pm$0.01}   &&\multicolumn {3}{c}{0.23$\pm$0.01} \\
$EW$(abs)~\AA                  &\multicolumn {3}{c}{1.9$\pm$0.4}     &&\multicolumn {3}{c}{1.7$\pm$0.4}\\
\hline
\end{tabular}

$^{\rm a}$in units 10$^{-14}$\ erg\ s$^{-1}$cm$^{-2}$.
\end{table*}


The emission line fluxes were measured using  Gaussian profile fitting. 
The errors of the line flux measurements include the errors
in the fitting of profiles and those in the placement of the continuum. 
They have been propagated in the
calculations of the elemental abundance errors.
The observed ($F$($\lambda$)) and corrected 
($I$($\lambda$)) emission line fluxes relative to the H$\beta$  
fluxes, the equivalent widths $EW$ of the emission lines,
the observed fluxes of H$\beta$, and the equivalent widths of the hydrogen 
absorption lines are listed in Table \ref{t:Intens}.

The H$\alpha$-to-H$\beta$ flux ratios in both H {\sc ii} regions are 
lower than the theoretical value (e.g., Brocklehurst 1971).
This is likely not due to data reduction problems as
the H$\alpha$-to-H$\beta$ flux ratios in other galaxies observed
during the same night are greater than the theoretical ones.
Therefore, an extinction coefficient $C$(H$\beta$)=0 was assumed  
for these \ion{H}{ii} regions and the emission-line fluxes were corrected
for Balmer line absorption only. 

\begin{figure}[hbtp]
    \psfig{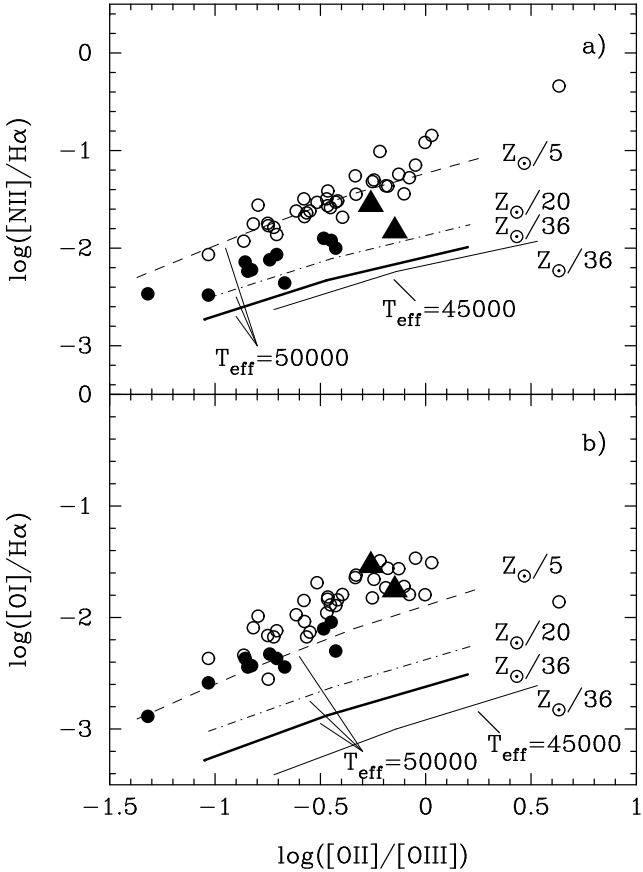}
    \caption{Dependence on a logarithmic scale of the line flux ratios ({\bf a}) 
log([N {\sc ii}]$\lambda$6584/H$\alpha$) and ({\bf b}) 
log([O {\sc i}]$\lambda$6300/H$\alpha$) on the 
log([O {\sc ii}]$\lambda$3727/[O {\sc iii}]$\lambda$5007) ratio.
Data for  galaxies with 12 + log(O/H) 
$\leq$ 7.6 and 12+log(O/H) $\geq$ 7.9 are shown by filled  and 
open circles, respectively. The positions of the brightest 
regions {\it a} and {\it b} in SBS 1129+576 are shown  by triangles. 
Thin and thick solid lines represent theoretical predictions from
photoionization models calculated with the CLOUDY code 
for ionizing stars with effective temperatures  
$T_{\rm eff}$ = 45 000 K and $T_{\rm eff}$ = 50 000 K and a heavy element
mass fraction $Z$ = $Z_\odot$/36. Dot-dashed and dashed lines correspond
to models with $T_{\rm eff}$ = 50 000 K and a heavy
element mass fractions  $Z$ = $Z_\odot$/20 and $Z_\odot$/5, respectively.
The ionization parameter decreases from log $U$ = --2.0 
(left) to log $U$ = --3.0 (right).
 }
    \label{fig:mart2}
\end{figure}

Some  diagnostic lines were studied to check for possible deviations 
of the H {\sc ii} region emission in SBS 1129+576 from the predictions 
of  photoionization models. 
For this purpose, data for the emission lines 
were collected for the galaxies from Izotov et al. 
(\cite{ITL94}, \cite{Izotov97}), Thuan et al. (\cite{til95}), 
Izotov \& Thuan (\cite{IT98a}),
Fricke et al. (\cite{Fricke00}), Noeske et al. (\cite{Noeske00}), 
and Guseva et al. (\cite{Guseva2001}). 
In total we use data for 46 star-forming galaxies: 
11 lowest-metallicity galaxies with an oxygen 
abundance 12 + log(O/H) $\leq$ 7.6 and 35 higher-metallicity galaxies with 
an oxygen abundance 12 + log(O/H) $\geq$ 7.9. 
Various emission line flux ratios 
are shown in Fig.~\ref{fig:mart2}. 


In this figure,
galaxies with low and high oxygen abundances split into two 
sequences, shown by filled and open circles, respectively. 
This separation is in overall agreement with photoionization H {\sc ii} 
region models.
However, regions {\it a} and {\it b} of SBS 1129+576 
(filled triangles in Fig.~\ref{fig:mart2}) 
having lower ionization parameters, 
lie outside the main location of low-metallicity galaxies. 
The line ratios 
[S {\sc ii}]$\lambda$6717+6731/H$\alpha$ and 
[N {\sc ii}]$\lambda$6584/H$\alpha$ in SBS 1129+576 are $\sim$ 3 times and
[O {\sc i}]$\lambda$6300/H$\alpha$  $\sim$ 5 times higher than the ones
in the sample of the 11 lowest-metallicity galaxies. 
The locations of regions {\it a} and {\it b} in Fig.~\ref{fig:mart2}
also significantly deviate from the model predictions calculated
with the CLOUDY code (Ferland \cite{F96}; Ferland et al. \cite{F98}) 
for a heavy element mass fraction $Z$ = $Z_\odot$/36 derived for region {\it a}
and for ionizing stars with effective temperatures of 
$T_{\rm eff}$ = 45 000 K (thin solid lines) and 
$T_{\rm eff}$ = 50 000 K (thick solid lines). 
By dot-dashed and dashed lines we also show the model predictions for 
$T_{\rm eff}$ = 50 000 K and two 
heavy element mass fractions of $Z_\odot$/20 and $Z_\odot$/5, respectively.
 
 The discrepancies between photoionization models and 
the observed line intensity ratios
(essentially the high [O {\sc i}]$\lambda$6300/H$\beta$ and
[S {\sc ii}]$\lambda$6717+6731/H$\beta$) are usually explained by some contribution 
of shock waves. However, Stasi\'nska \& Izotov (\cite{StasinskaIz02}) from the analysis of 
a large sample ($\sim$ 400) of  {\sc H ii} galaxies proposed another model to explain 
the enhancement of these lines in low-metallicity {\sc H ii} regions
by invoking 
chemical inhomogeneities and self-enrichment by the heavy elements.
Their models take into account the time evolution of an ionizing cluster in 
the simple case of an expanding bubble. 
The models predict an increase of the  [O {\sc i}]$\lambda$6300/H$\beta$
ratio in better agreement with observations of low-metallicity {\sc H ii} regions.
Nevertheless, the hypothesis of shock heating may also be considered 
as an alternative explanation.

A two-zone photoionized H {\sc ii} region model has been assumed for the
abundance determination. The electron temperature $T_{\rm e}$(O {\sc iii}) in 
the high-ionization region has been derived from the observed flux ratio 
[O {\sc iii}]$\lambda$4363/($\lambda$4959+5007), 
using a five-level atom model (Aller \cite{Aller84}) 
with atomic data from Mendoza (\cite{Mendoza83}). 
The electron temperature $T_{\rm e}$(O {\sc ii})
in the low-ionization region has been obtained using the empirical relation 
between $T_{\rm e}$(O {\sc ii}) and $T_{\rm e}$(O {\sc iii}) from the  
H {\sc ii} region photoionization models by Stasi\'nska (\cite{Stasinska90}).
The [S {\sc ii}]$\lambda$6717/$\lambda$6731 ratio was used to derive the 
electron number density $N_{\rm e}$(S {\sc ii}). 

The abundances of O$^{+2}$, Ne$^{+2}$ and He$^+$
were derived applying the electron temperature $T_{\rm e}$(\ion{O}{iii}).
The electron temperature $T_{\rm e}$(\ion{O}{ii}) is adopted for the O$^+$ and
N$^+$ ionic abundance determination.

The total oxygen abundance is 
the sum of the O$^+$ and O$^{+2}$ abundances. 
The total abundances of other heavy elements were derived using ionization correction 
factors following Izotov et al. (\cite{ITL94}, \cite{Izotov97}) and 
Thuan et al. (\cite{til95}). 
The ionic and heavy element abundances for regions {\it a} and {\it b}  
 together with electron temperatures 
and electron number densities are given in Table \ref{t:Chem} along with 
the adopted ionization correction factors (ICF). 

The oxygen abundances  
12 + log(O/H) = 7.36 $\pm$ 0.10 ($Z$=$Z_\odot$/36) and 7.48 $\pm$ 0.12 
($Z$=$Z_\odot$/28) for regions {\it a} and {\it b}, respectively, are slightly
different but still consistent within the 1$\sigma$ errors.
  Note that the oxygen abundance in region {\it a} may be underestimated due to the 
additional contribution of shock enhancement of the [O {\sc iii}]$\lambda$4363 
emission line. Such an effect can be present in region {\it b}, but
it is expected to be much larger in region {\it a} with the much weaker 
[O {\sc iii}]$\lambda$4363 emission line (Table~\ref{t:Intens}). 

The abundance ratios of nitrogen and neon to oxygen (Table \ref{t:Chem}) are in
agreement with those derived in other lowest-metallicity dwarf galaxies 
(Izotov \& Thuan \cite{IT99}).

The low metallicity of SBS 1129+576 makes it potentially suitable for 
He abundance determination. For this, we use the three strongest 
He {\sc i} $\lambda$4471, $\lambda$5876 and $\lambda$6678 emission lines. 
The helium abundance derived from the corrected fluxes 
of all observed He {\sc i} emission lines is shown in Table \ref{t:Chem}. The mean 
values for the He mass fraction of $Y$ = 0.220 $\pm$ 0.027 
(region {\it a}) and $Y$ = 0.202 $\pm$ 0.042 (region {\it b}), 
though formally consistent, are significantly lower 
than the primordial 
He mass fraction $Y_{\rm p}$ = 0.244 $\pm$ 0.002, derived by extrapolating the 
$Y$ vs O/H linear regression to O/H = 0 for the sample of 45 low-metallicity
dwarf galaxies (Izotov \& Thuan \cite{IT98a}), or to $Y_{\rm p}$ = 0.245 $\pm$ 0.002 
derived for the most metal-deficient BCDs I Zw 18 and SBS 0335--052 
(Izotov et al. \cite{ICFGGT99}). 

The likely source of lower values of $Y$ in SBS 1129+576
is underlying stellar He {\sc i} absorption superposed on the He~{\sc i}
emission lines. Comparing the equivalent widths of He~{\sc i} emission
lines (Table \ref{t:Intens}) with those predicted by population
synthesis models for He {\sc i} absorption lines
(e.g., Gonz\'alez Delgado et al. 1999) we conclude that 
underlying stellar absorption can decrease the fluxes of the
He {\sc i} emission lines by 
as much as $\sim$ 10 -- 20\% for He {\sc i} $\lambda$5876 (assuming the
equivalent widths of He {\sc i} $\lambda$5876 and 
He {\sc i} $\lambda$4471 absorption lines
to be similar) and more for other  He {\sc i}  lines.
Hence, despite its low metallicity SBS 1129+576 is not a good candidate 
for primordial helium abundance determination because
of the large effect of the underlying stellar absorption.
\begin{table}[tbh]
\caption{Element abundances in regions {\it a} and {\it b}.}
\label{t:Chem}
\begin{tabular}{lccc} \hline \hline
Value                               & region {\it a}      && region {\it b}  \\ \hline
$T_{\rm e}$(O {\sc iii})(K)               &18930$\pm$2540 && 20560$\pm$2660 \\
$T_{\rm e}$(O {\sc ii})(K)                &15350$\pm$1930 && 15850$\pm$2180 \\
$T_{\rm e}$(S {\sc iii})(K)               &17410$\pm$2110 && 18760$\pm$2210 \\
$N_{\rm e}$(S {\sc ii})(cm$^{-3}$)      &   40$\pm$120  &&    10$\pm$10   \\ \\
O$^+$/H$^+$($\times$10$^5$)         &1.092$\pm$0.351&& 1.298$\pm$0.594\\
O$^{+2}$/H$^+$($\times$10$^5$)      &1.215$\pm$0.378&& 1.691$\pm$0.907\\
O/H($\times$10$^5$)                 &2.307$\pm$0.516&& 2.989$\pm$1.084\\
12 + log(O/H)                       &7.363$\pm$0.097&& 7.476$\pm$0.121\\ \\
N$^{+}$/H$^+$($\times$10$^7$)       &2.785$\pm$0.896&& 5.699$\pm$2.021\\
ICF(N)$^{\rm a}$                    &2.11\,~~~~~~~~~&& 2.30\,~~~~~~~~~~\\
log(N/O)                            &--1.593$\pm$0.241~~&&--1.414$\pm$0.298~~\\ \\
Ne$^{+2}$/H$^+$($\times$10$^5$)     &0.213$\pm$0.070&& 0.320$\pm$0.165\\
ICF(Ne)$^{\rm a}$                   &1.90\,~~~~~~~~~&&1.77\,~~~~~~~~~~\\
log(Ne/O)                           &--0.756$\pm$0.199~~&&--0.722$\pm$0.245~~\\ \\
He$^+$/H$^+$($\lambda$4471)         &0.0686$\pm$0.0313&& 0.0819$\pm$0.0399\\
He$^+$/H$^+$($\lambda$5876)         &0.0684$\pm$0.0094&& 0.0603$\pm$0.0142\\
He$^+$/H$^+$($\lambda$6678)         &0.0868$\pm$0.0251&& 0.0692$\pm$0.0447\\
He$^+$/H$^+$                        &                 &&                  \\
(weighted mean)                     &0.0705$\pm$0.0085&& 0.0633$\pm$0.0129\\
He/H                                &0.0705$\pm$0.0085&& 0.0633$\pm$0.0129\\
$Y$                                 &0.2199$\pm$0.0271&& 0.2019$\pm$0.0418\\ \hline
\end{tabular}

$^{\rm a}$ICF is the ionization correction factor.
\end{table}

\section{Age of the underlying stellar population \label{age}}

The low metallicity and relatively blue $(V-I)$ colour of SBS 1129+576
make this object a good young galaxy candidate.
In this section we consider the properties of the stellar populations
in the galaxy and discuss its evolutionary status.
The observed properties of stellar populations are dependent on the 
metallicity and star formation history. They also can be influenced
by interstellar extinction and emission of the ionized gas. 
Therefore,
we use all available spectroscopic and photometric observational data 
on 5 LSB regions  
to put consistent constraints on the age of stellar populations in SBS 1129+576.

Emission and absorption hydrogen Balmer
lines are seen in the spectra of a large part of the galaxy along the slit.
This allows us to study the age of stellar populations, using two methods,  
based on the time evolution of equivalent widths of (1) nebular emission 
Balmer lines and (2) stellar absorption Balmer lines.
The advantage of these methods is that they are extinction-insensitive.
This is very important because the only way to derive 
interstellar extinction from optical spectra is to compare observed and
theoretical decrements of Balmer emission lines. However, in the extended
low-intensity regions the emission lines are weak or absent, making estimates
of the interstellar extinction uncertain.

The age of the stellar population in a galaxy can be obtained from 
a third method, comparing the observed and theoretical spectral energy 
distributions (SED), the latter computed with various ages and histories of 
star formation for the stellar population. 
However, the shape of the observed continuum is dependent on both 
age and interstellar extinction. 
If no other information on the stellar population is available
(e.g., the age derived from hydrogen equivalent widths), the 
extinction coefficient is set equal to zero (for a spectrum with no Balmer 
emission lines), and the age can simply be inferred by fitting the 
observed spectrum with theoretical SEDs. 
This method gives a maximum age  among possible age estimates.
If, on the other hand, the ages of stellar populations 
can be estimated by some other methods, then the reddening can be derived from 
the SEDs. 

Finally, to study stellar populations in SBS 1129+576, we use broad-band
photometry.
However, similar to the SED fitting method, 
extinction also affects age determination  based on photometric data.
Additionally, ionized gas emission can significantly influence 
both the observed SEDs and broad-band colours.


\begin{figure}[hbtp]
    \psfig{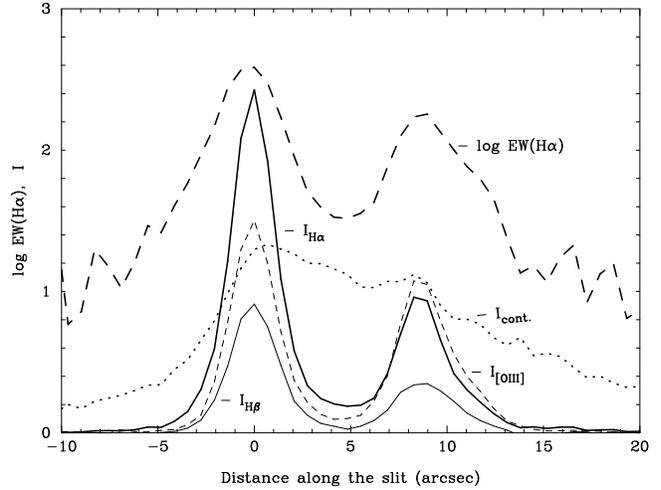}
    \caption{Distributions of the H$\alpha$ equivalent width on a logarithmic 
     scale (thick dashed line), and fluxes of the H$\alpha$ (thick solid line),
     H$\beta$ (thin solid line), [O {\sc iii}] $\lambda$5007 (thin dashed
     line) emission lines and the continuum near H$\beta$ (dotted line)  
     along the major axis of SBS 1129+576. 
     The axis origin is set to the position of region {\it a}
     (see Fig. \ref{f1}). The fluxes are in arbitrary units.
     } 
    \label{fig:ew}
\end{figure}

Before dealing with stellar populations, we first consider how important  
is the contribution of ionized gas emission to the total light.
The variations of the equivalent width of the H$\alpha$ emission line
and the fluxes of the strongest emission lines along the major axis of 
SBS 1129+576 are shown in Fig.~\ref{fig:ew}. 
The maximum  $EW$(H$\alpha$) = 385\AA\ and  
$EW$(H$\beta$) = 64 \AA\ are derived in region {\it a}.
The lower $EW$(H$\alpha$) and $EW$(H$\beta$) for
these regions listed in Table ~\ref{t:Intens} are 
due to different apertures used, of 5\arcsec$\times$2\arcsec\ 
(Table \ref{t:Intens}) 
and of 0\farcs69$\times$2\arcsec\ (Fig. \ref{fig:ew}).
The H$\alpha$ equivalent widths in other regions are much lower. 
Hence,  the contribution of gaseous emission to broad-band fluxes is small in the  bright 
H {\sc ii} regions  and negligible in the LSB component. 

\subsection{Age from Balmer nebular emission lines\label{under_1}}

The largest equivalent widths $EW$(H$\alpha$) are measured in regions {\it a} and
{\it b} (Fig. \ref{fig:ew}), implying that the light from these regions
is dominated by young stellar populations. 
The equivalent widths of H$\alpha$, H$\beta$ emission lines 
in LSB regions are much smaller, implying a higher contribution to the light
of the old stellar populations. 
The fluxes and 
equivalent widths of the H$\alpha$, H$\beta$ (and H$\gamma$
where possible) emission lines were measured in the spectra of five 
LSB regions   and listed together with 
errors in Table~\ref{t:emhahb}.
Because the H$\beta$ emission line is narrower than the absorption line 
and does not fill the absorption component, its flux was measured
using the continuum level at the bottom of the absorption line.
This level has been chosen by visually interpolating from the absorption line
wings to the center of the line.

\begin{figure}[hbtp]
    \psfig{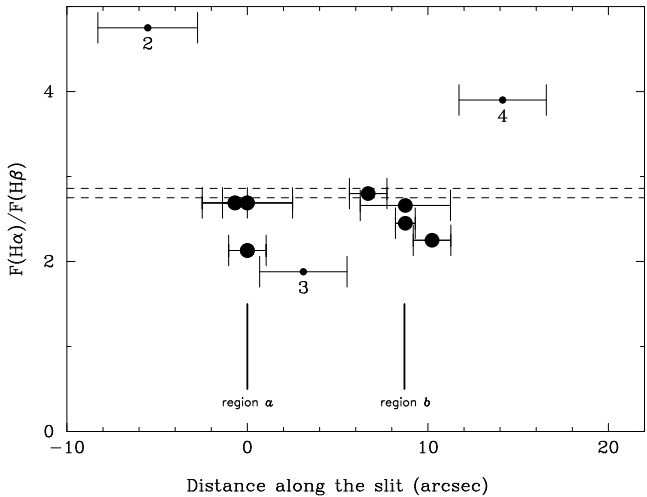}
    \caption{The observed H$\alpha$/H$\beta$ emission line 
     flux ratios 
     in the different regions of SBS 1129+576.
     Solid horizontal bars indicate the apertures
     wherein the spectra
     were extracted. 
     Large  circles show
     the H$\alpha$/H$\beta$ flux ratios for regions 
     {\it a} and {\it b}, computed within different apertures.
     The position of the centers of the two brightest H {\sc ii} regions is
     marked by vertical solid lines.
     Small circles show the same ratios for the fainter 
     regions 2, 3 and 4.
     The upper and lower dashed horizontal lines denote the theoretical 
     $F$(H$\alpha$)/$F$(H$\beta$) flux ratios for ionization-bounded (case B)  
     \ion{H}{ii} regions
     with $T_{\rm e}$ = $10^4$~K and  $T_{\rm e}$ = 2 $\times$ $10^4$~K,
     respectively.
     } 
    \label{fig:ratha_hb}
\end{figure}

 The extinction cannot be derived from the Balmer decrement because 
the observed H$\alpha$-to-H$\beta$ emission line flux 
ratios for most of the selected regions
are lower than theoretical predictions of ionization-bounded
H {\sc ii} region models (Tables~\ref{t:Intens}, \ref{t:emhahb} 
and Fig.~\ref{fig:ratha_hb}).
Therefore an extinction coefficient $C$(H$\beta$) of zero
is adopted.
Two dashed lines in Fig.~\ref{fig:ratha_hb} denote the theoretical 
values  $F$(H$\alpha$)/$F$(H$\beta$) = 2.86
for $T_{\rm e}$ = $10^4$~K and  2.75 for
$T_{\rm e}$ = 2 $\times$ $10^4$~K. H$\alpha$ and H$\beta$ fluxes have been 
measured in each  region except for the  outermost regions 1 and 5 
where only  H$\alpha$ emission is present. Only in the two regions, 2 and 4, is the 
H$\alpha$/H$\beta$ flux ratio larger than the theoretical 
recombination  flux ratio.
No correction for the absorption line equivalent width 
has been made in these regions.
 The extinction coefficients $C$(H$\beta$) for 
regions 2 and 4 are shown in Table \ref{t:emhahb}. 
Note that $C$(H$\beta$) for region 4 is very uncertain, because of the very 
low equivalent width $EW$(H$\beta$), which is comparable to the $EW$
of the H$\beta$ absorption line.


\begin{figure}[hbtp]
    \psfig{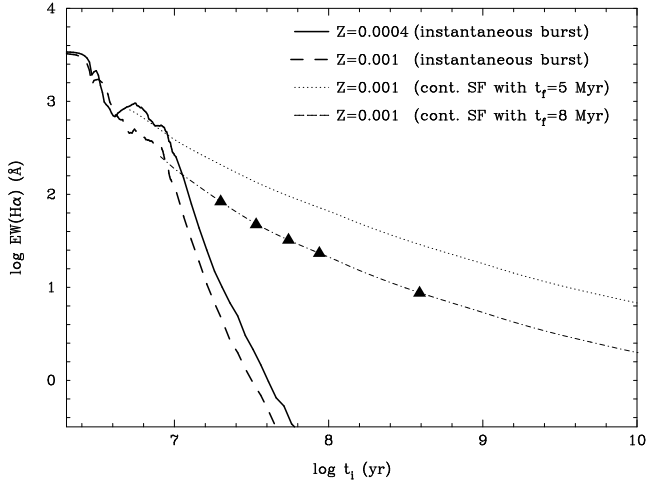}
    \caption{Time evolution of the equivalent widths 
    of nebular
    emission line H$\alpha$ 
    for an instantaneous burst with a heavy element mass fraction 
    $Z$ = $Z_\odot$/20  (dashed line) and
    $Z$ = $Z_\odot$/50  (solid line).
    Model predictions are also shown for the case of 
    constant continuous star formation starting at an age defined by the 
    abscissa $t_{\rm i}$ and stopping at $t_{\rm f}$, with $t_{\rm f}$ = 5 Myr
    (dotted line) and $t_{\rm f}$ = 8 Myr (dash-dotted line), both
    with $Z$ = $Z_\odot$/20. 
    Triangles 
    show the positions of LSB regions with the observed 
    $EW$(H$\alpha$) 
    superposed  on the modeled curve   
    for continuous star formation stopping at $t_{\rm f}$ = 8 Myr.
     } 
    \label{fig:ewhbha}
\end{figure}

The dependence of the H$\alpha$ emission line equivalent width
on age is shown in Fig.~\ref{fig:ewhbha}. 
Here we assume ionization-bounded  \ion{H}{ii} region model.
The model H$\alpha$ equivalent widths in the case of density-bounded 
\ion{H}{ii} region are lower than those in Fig.~\ref{fig:ewhbha}.
Since the temporal evolution of emission 
line equivalent widths depends on the star formation history we consider two
limiting cases: an instantaneous burst model and models with continuous
star formation. The instantaneous burst model equivalent widths 
of H$\alpha$ and H$\beta$ emission lines
are calculated
using the galactic evolution code PEGASE.2 (Fioc \& Rocca-Volmerange 
\cite{F97}). In Fig.~\ref{fig:ewhbha} the $EW$(H$\alpha$) 
for the heavy element mass 
fraction $Z_\odot$/50 is shown by a solid line and that for the heavy 
element mass 
fraction $Z_\odot$/20 by a dashed line. PEGASE.2 is based on the 
Padua stellar evolutionary models (Bertelli 
et al. \cite{Bertelli94}) and stellar atmosphere models from Lejeune et al. 
(\cite{Lejeune98}). 
An initial mass function with a Salpeter 
slope ($\alpha$ = --2.35), and upper and lower mass limits of 120 
$M_\odot$ and 0.1 $M_\odot$ are adopted for all our calculations with
the PEGASE.2 code.

An instantaneous burst model is most appropriate for 
the star-forming regions {\it a} and {\it b}. 
The maximum equivalent widths of Balmer emission lines in region {\it a} 
($EW$(H$\alpha$) = 385\AA\ and $EW$(H$\beta$) = 64 \AA) are consistent
with an instantaneous burst age of 6 Myr for a heavy element mass fraction 
of $Z$ = $Z_\odot$/20 and  of 9 Myr for $Z$ = $Z_\odot$/50 
(Fig.~\ref{fig:ewhbha}). The maximum value $EW$(H$\alpha$) = 180\AA\ 
for region {\it b} is consistent with an instantaneous burst age of 9 
Myr for $Z$ = $Z_\odot$/20 and 11 Myr for $Z$ = $Z_\odot$/50.
The $EW$s of Balmer emission lines in regions {\it a} and {\it b} are probably
slightly underestimated due to the significant ($\ga$ 30\%; Fig. \ref{f2}a) 
line-of-sight contribution of the emission from the LSB component
at the respective position.
Given the steep decrease of the $EW$(H$\alpha$) as a function of time for
7 $\la$ log($t$ yr) $\la$ 7.4 (Fig. \ref{fig:ewhbha}) this will not 
significantly reduce the instantaneous burst age for regions {\it a} and {\it b}.

In continuous star formation models we adopt a constant star 
formation rate in the interval between the 
time $t_{\rm i}$ when star formation starts and $t_{\rm f}$ when it stops.
Time is zero now and increases to the past. 

We use the model equivalent widths of hydrogen emission 
lines and SEDs for instantaneous bursts (Fioc \& 
Rocca-Volmerange \cite{F97})
to calculate the temporal evolution of the equivalent widths of hydrogen 
emission lines in the case of continuous star formation with the constant SFR. 
The results are given in Fig. \ref{fig:ewhbha} for a heavy element mass
fraction $Z_{\odot}$/20.
The temporal dependence of the equivalent width of the H$\alpha$   
emission line is shown for continuous star
formation starting at time $t_{\rm i}$, as defined by the abscissa value, and 
stopping at $t_{\rm f}$ = 5 Myr (dotted line) and $t_{\rm f}$ = 8 Myr
(dash-dotted line). 
The equivalent width of the H$\alpha$ emission
line in the spectrum of a stellar population formed between $t_{\rm i}$ 
and $t_{\rm f}$ is the value 
of $EW$ at time $t_{\rm i}$ (Fig. \ref{fig:ewhbha}). 
At a fixed $EW$, the younger are the 
youngest stars, the larger is the time interval $t_{\rm i} - t_{\rm f}$, 
i.e. the older are the oldest stars.

The models of the continuous star formation apply best to the
extended stellar component in SBS 1129+576.
The observed $EW$(H$\alpha$)  for regions 1 -- 5 
(Table \ref{t:emhahb}) 
are shown in Fig.~\ref{fig:ewhbha} by triangles
 on the theoretical curve with the
star formation stopping at $t_{\rm f}$ = 8 Myr ago and correspond to ages
$t_{\rm i}$ in the range 20 -- 400 Myr.
If instead the star formation 
continues until $t_{\rm f}$ = 5 Myr (dotted line), then the age $t_{\rm i}$ of
the oldest stars, as derived from the equivalent widths of the H$\alpha$ and
H$\beta$ emission lines, should be in the range 50 Myr -- 5 Gyr. 
However, the latter star formation history seems to be less likely if a
stellar IMF with upper mass cut-off of $M_{\rm}$ = 100 $M_\odot$ is assumed.
This is because massive stars with  $M$ = 40 $M_\odot$
are expected to be present in region with  $t_{\rm f}$ = 5 Myr while
the H$\alpha$ luminosities of regions 1 -- 5 
correspond to  fluxes of ionizing photons lower than that produced by 
single massive star with $M$ = 40 -- 60 $M_\odot$.
 For continuous star formation with  $t_{\rm f}$ = 0 the upper mass cut-off 
of the stellar IMF must be even lower.


\begin{table*}[tbh]
\caption{Fluxes, equivalent widths of the H$\alpha$, H$\beta$ and H$\gamma$ 
emission lines and the extinction coefficient $C$(H$\beta$) 
in LSB regions.}
\label{t:emhahb}
\begin{tabular}{lccrrrrrrr} \hline \hline
Region &\multicolumn{1}{c}{Distance$^{\rm a}$}
& \multicolumn{1}{c}{Aperture$^{\rm b}$}
& \multicolumn{1}{c}{$F$(H$\alpha$)$^{\rm c}$}
  &\multicolumn{1}{c}{$EW$(H$\alpha$)$^{\rm d}$} 
&\multicolumn{1}{c}{$F$(H$\beta$)$^{\rm c}$}  
&\multicolumn{1}{c}{$EW$(H$\beta$)$^{\rm d}$} 
&\multicolumn{1}{c}{$F$(H$\gamma$)$^{\rm c}$} 
&\multicolumn{1}{c}{$EW$(H$\gamma$)$^{\rm d}$} 
&\multicolumn{1}{c}{$C$(H$\beta$)}
   \\ \hline

1 &--7.9~\, &2.0$\times$6.2 & 4.2$\pm$0.5 & 32.3$\pm$2.5 & ...~~~~      &   ...~~~~       &   ...~~~~      &   ...~~~~       &   ...~~~~~       \\
2 &--5.5~\, &2.0$\times$5.5 & 18.4$\pm$0.6 & 83.5$\pm$1.8 & 3.9$\pm$0.5 & 9.6$\pm$0.7 &    ...~~~~      &   ...~~~~       & 0.33$\pm$0.07  \\
3 &  3.1 &2.0$\times$4.8 & 25.4$\pm$0.6 & 47.3$\pm$0.8 & 13.5$\pm$0.6 & 16.2$\pm$0.5 & 5.6$\pm$0.5 & 4.2$\pm$0.3 &   ...~~~~~       \\
4 & 14.1~\, &2.0$\times$4.8 & 4.1$\pm$0.3 & 23.3$\pm$1.9 & 1.1$\pm$0.3 & 2.4$\pm$0.7 &   ...~~~~      &   ...~~~~       & 0.20$\pm$0.06  \\ 
5 & 17.9~\, &2.0$\times$4.8 & 0.9$\pm$0.3 & 8.7$\pm$2.5 & ...~~~~      &  ...~~~~       &  ...~~~~      &   ...~~~~       & ...~~~~~       \\

\hline
\end{tabular}

$^{\rm a}$distance in arcsec from region {\it a}. Negative and positive values correspond to regions located 
respectively to the southeast and northwest from region {\it a}. \\
$^{\rm b}$aperture $x$ $\times$ $y$ where $x$ is the slit width and $y$ the 
size along the slit in arcsec. \\
$^{\rm c}$in units 10$^{-16}$\ erg\ s$^{-1}$cm$^{-2}$. \\
$^{\rm d}$in \AA. \\
\end{table*}

The small number of massive stars and hence the
stochastic nature of the IMF by 
region  can influence our
age estimates from hydrogen emission line equivalent widths. 
From the observed  fluxes of the 
H$\beta$ emission line for regions 1 -- 5
we find that 3 -- 23 O7{\sc v} or 35 -- 233 B0{\sc v} stars are required to 
ionize the gas 
(see Table \ref{t:emhahb}). Here,  Lyc photon fluxes of 
10$^{49}$ s$^{-1}$ and 10$^{48}$ s$^{-1}$ were respectively adopted for a 
single O7{\sc v} and B0{\sc v} star (Vacca et al. \cite{Vacca96}).
Then the total stellar mass of the single stellar population was estimated 
to be in the range 10$^4$ -- 10$^5$ $M_\odot$ assuming an IMF with Salpeter 
slope and 
upper and lower stellar mass limits of 120 $M_\odot$ and 2 $M_\odot$.
The total mass  for the different regions is in the same range as in 
the case of SBS 0940+544 (Guseva et al. \cite{Guseva2001}).
  Cervi\~no et al. (\cite{Cervino00}) find that
in the range of equivalent widths $EW$(H$\beta$) = 1 -- 10 \AA\  
 which is typical for regions 1--5 
(see Table \ref{t:emhahb}), the dispersion
of age at fixed $EW$(H$\beta$) is not greater than 5 -- 10\% if the total
 mass of the cluster lies in the range 10$^4$ -- 10$^5$ $M_\odot$.
 Hence, we conclude that our age estimate is not significantly affected
by stochastic effects.

\subsection{Age from the Balmer stellar absorption lines\label{under_2}}

We also use a second extinction-insensitive
method to estimate stellar population ages from the
equivalent widths of Balmer absorption lines  H$\gamma$ and H$\delta$.
Other higher-order hydrogen Balmer absorption lines have not been used for 
age determination due to the relatively noisy spectrum at short wavelengths.
The ages were derived from the calibration of equivalent widths of absorption 
Balmer lines versus age calculated by Gonz\'alez Delgado et al. 
(\cite{GonLeith99b}) {for ages $\la$ 1 Gyr}. 
Their models predict in the case of an instantaneous burst of 
star formation a steady increase of the equivalent widths with age
for ages ranging from 1 Myr to 1 Gyr. However, for ages $\ga$ 1 Gyr the 
situation is the opposite and 
the equivalent widths of the absorption lines decrease with time (Bica \& Alloin
 1986).



\begin{table*}[tbh]
\caption{Equivalent widths of H$\gamma$ and H$\delta$ absorption lines in LSB regions.}
\label{t:abshdhg}
\begin{tabular}{lcccccc} \hline \hline
Region & Distance$^{\rm a}$& Aperture$^{\rm b}$&
$EW$(H$\delta$)$^{\rm c}$ &$EW$(H$\gamma$)$^{\rm c}$  
&$EW$(H$\delta$)$_{\rm cor}$$^{\rm c}$  
&$EW$(H$\gamma$)$_{\rm cor}$$^{\rm c}$   \\ \hline

1 & --7.9~\, &2.0$\times$6.2 & 5.6 $\pm$1.1 & 5.3 $\pm$1.3 & 5.6 $\pm$1.1 & 5.3 $\pm$1.3 \\    
2 & --5.5~\, &2.0$\times$5.5 & 5.7 $\pm$0.6 &        ...       & 6.4 $\pm$0.7 &        ...    \\
3 &  3.1 &2.0$\times$4.8 & 4.7 $\pm$0.4 &        ...       & 7.1 $\pm$0.5 &        ...    \\  
4 & 14.1\, &2.0$\times$4.8 & 6.9 $\pm$0.5 & 4.1 $\pm$0.5 & 7.1 $\pm$0.8 & 4.6 $\pm$0.5 \\
5 & 17.9\, &2.0$\times$4.8 & 5.7 $\pm$0.7 & 6.5 $\pm$0.9 & 5.7 $\pm$0.7 & 6.5 $\pm$0.9 \\ 
\hline
\end{tabular}

$^{\rm a}$distance in arcsec from the brightest H {\sc ii} region {\it a}.  Negative and 
positive values correspond to regions located 
respectively to southeast and northwest from region {\it a}. \\
$^{\rm b}$aperture $x$ $\times$ $y$ where $x$ is the slit width and $y$ the 
 size along the slit in arcsec. \\
$^{\rm c}$in \AA.  \\
\end{table*}

The hydrogen absorption lines, created by the 
underlying stellar populations, are seen in all positions along the 
slit.
 In two regions, 2 and 3,   
the H$\gamma$ absorption line was not used for the equivalent width
measurements because the contamination by the
H$\gamma$ and [O {\sc iii}]$\lambda$4363 emission lines is too strong 
(Figs. ~\ref{fig:spfit_c1}, ~\ref{fig:spfit_c2}).

\begin{figure}[hbtp]
    \psfig{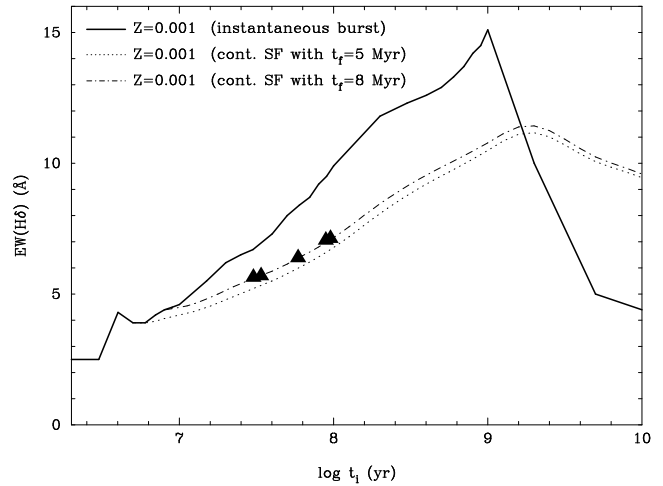}
    \caption{Dependence of the equivalent width of 
    H$\delta$ absorption line for an instantaneous burst 
    on age for ages $>$ 1 Gyr from Bica \& Alloin (1986)  and for ages $\leq$ 1 Gyr
       with metallicity $Z$=$Z_\odot$/20  from Gonz\'alez Delgado et al. (1999)
     is shown by solid line.
    Model predictions are also shown for the case of 
    constant continuous star formation starting at an age defined by the 
    abscissa $t_{\rm i}$ and stopping at $t_{\rm f}$, with  $t_{\rm f}$ = 5 Myr
    (dotted line) and $t_{\rm f}$ = 8 Myr (dash-dotted line), both with
    $Z$ = $Z_\odot$/20.
    Triangles show the positions of the corrected 
    $EW$(H$\delta$) for LSB regions 1--5
    superposed on the models of continuous
    star formation with $t_{\rm f}$ = 8 Myr.  
         } 
    \label{fig:abshghd}
\end{figure}

The measured equivalent widths of absorption lines were corrected for the 
contribution of the nebular emission in the same way as in 
Guseva et al. (\cite{Guseva2001}).
No extinction correction has been applied to emission lines superposed on 
absorption profiles.
Table~\ref{t:abshdhg} lists the uncorrected 
equivalent widths of 
the H$\gamma$ and H$\delta$ absorption lines with errors
and those corrected 
for the contribution of the nebular emission.
The corrected equivalent widths of H$\gamma$ and H$\delta$ do not
show statistically significant spatial variations.
Note that the corrections are larger for the brighter regions.
 A careful placement of the continuum is very important 
for deriving accurate $EW$s.
We choose points in the spectrum free of nebular 
and stellar lines for the determination of the continuum level. 
Then the continuum was fitted by cubic splines. The uncertainties of 
the continuum level were estimated from several different measurements of 
the equivalent widths of Balmer absorption lines with independent continuum 
fittings. These uncertainties are of the same order as the errors in 
Table~\ref{t:abshdhg}, obtained from the fitting of line profiles with Gaussians 
and from continuum noise.

The instantaneous burst dependence on the age 
of the equivalent width of 
the H$\delta$ absorption line (solid line)   
(Bica \& Alloin (1986) for ages $>$ 1 Gyr; and Gonz\'alez Delgado et al. 
(\cite{GonLeith99b}) 
for ages $\leq$ 1 Gyr at a metallicity $Z$ = $Z_\odot$/20) is 
shown in Fig.~\ref{fig:abshghd}.

The temporal evolution of the H$\gamma$ and H$\delta$ 
absorption line equivalent widths in the case of continuous star formation
is calculated similarly to that of the H$\alpha$ and H$\beta$ 
emission line equivalent widths described in Sect. \ref{under_1}.
More specifically, we use the model 
and empirical
equivalent widths of hydrogen 
absorption lines 
(Gonz\'alez Delgado et al. \cite{GonLeith99b};
Bica \& Alloin \cite{Bica86})
and SEDs for instantaneous bursts (Fioc \& 
Rocca-Volmerange \cite{F97}) to calculate the temporal evolution of the 
equivalent widths of hydrogen absorption lines in the case of continuous star 
formation with constant SFR. 
The results are given in Fig. \ref{fig:abshghd}. 
The temporal dependence of the equivalent width of the H$\delta$ 
absorption line is shown for star
formation starting at time $t_{\rm i}$, as defined by the abscissa value, and 
stopping at $t_{\rm f}$ = 5 Myr (dotted line) and $t_{\rm f}$ = 8 Myr
(dash-dotted line). 

The measured $EW$(H$\delta$) are shown in Fig. \ref{fig:abshghd}
on the theoretical curve with the star formation stopping at $t_{\rm f}$ = 
8 Myr ago and are marked by triangles. 
Their positions are consistent with an age $t_{\rm i}$ in the range of  
$\sim$ 30 -- 140 Myr. A similar age is obtained 
from the nebular emission lines. If instead the star formation is
continuing until $t_{\rm f}$ = 5 Myr (dotted lines), then the age of
the oldest stars, as derived from the equivalent width of the
H$\delta$ absorption line, differs significantly from that derived from the 
H$\alpha$ emission line.
Furthermore, the discrepancy between ages derived
from the equivalent widths of the emission and absorption lines is very
large for a model with $t_{\rm f}$ = 0.
Hence, from the comparison of hydrogen emission
and absorption line equivalent widths we exclude models of star formation
in which stars are continuously forming with a constant star formation rate 
between 0 and 10 Gyr.
Finally, we choose the case with $t_{\rm f}$ = 8 Myr  to be consistent 
with the age derived from
the equivalent widths of the H$\alpha$ emission line.

\subsection{Age from the spectral energy distribution \label{SED}}

The shape of the spectrum reflects both the properties of the
stellar population and the reddening.
Therefore  only a combination of the spectral energy distribution method with 
the methods discussed in Sect. \ref{under_1} and \ref{under_2} allows us to 
derive simultaneously age and interstellar extinction.

We use the galactic evolution code PEGASE.2 (Fioc \& 
Rocca-Volmerange \cite{F97}) to produce a grid of theoretical SEDs
for an instantaneous burst of star formation and ages ranging between  
0 and 10 Gyr, and a heavy element mass fraction of $Z$ = $Z_\odot$/20.
Because of the low equivalent widths of hydrogen emission lines in regions 
1 -- 5 the gaseous emission is not included in the spectral energy 
distributions. Thus, photometric and spectroscopic data give 
direct information about stellar populations when
interstellar extinction is known.

\begin{figure}[hbtp]
    \psfig{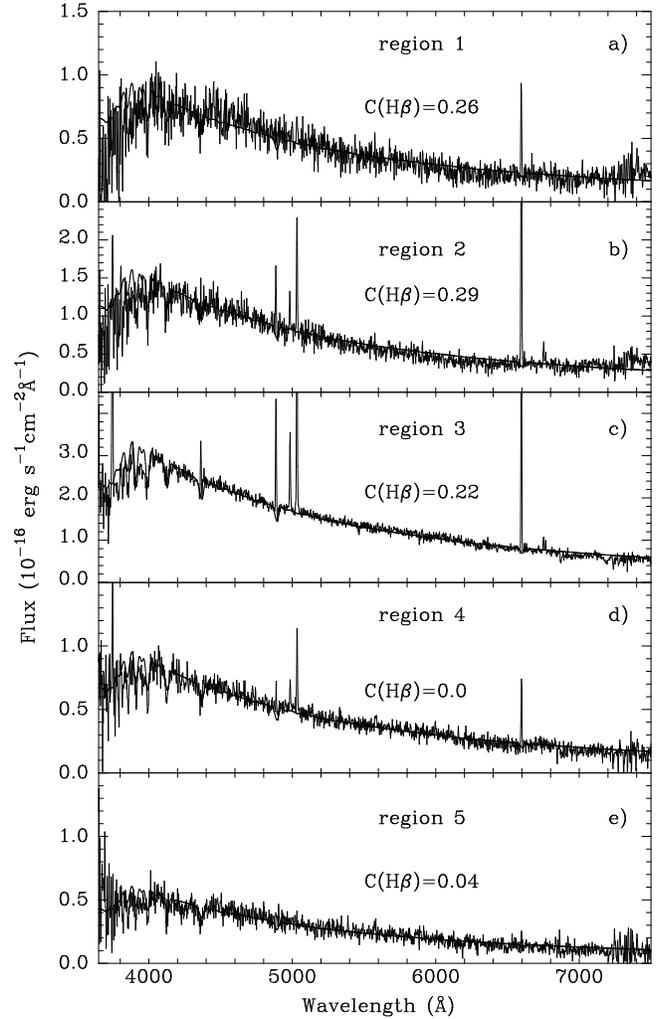}
    \caption{Spectra of regions 1 -- 5 on which 
     synthetic continuum spectral energy distributions are superposed. 
     Synthetic SEDs in ({\bf a}) -- ({\bf e}) are calculated for a stellar population 
     formed continuously with a constant star formation rate between 
     8 and 100 Myr ago. Extinction coefficients are derived from the best fits of 
     observed and calculated SEDs. Each spectrum is labeled by 
     its respective $C$(H$\beta$). 
     The SEDs are superposed on the spectra corrected for the extinction.
     }
    \label{fig:spfit_c1}
\end{figure}

\begin{figure}[hbtp]
    \psfig{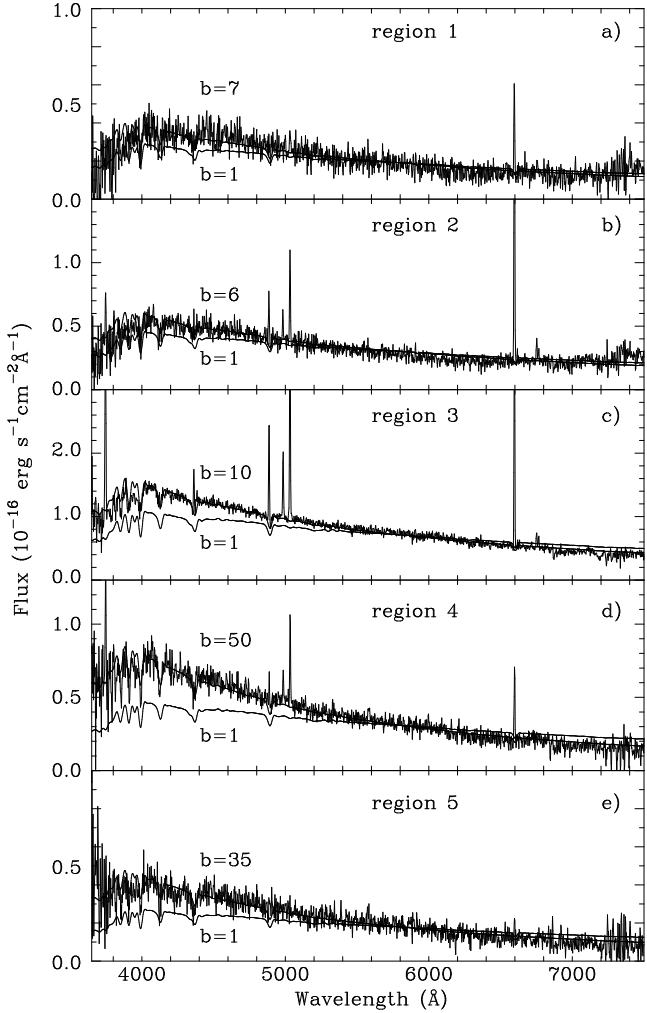}
    \caption{Spectra of regions 1 -- 5 on which synthetic continuum 
   spectral energy distributions are superposed. 
   Synthetic SEDs shown in ({\bf a}) -- ({\bf e}) and labeled by $b$ $\equiv$ 
   SFR($t$ $\leq$ 100 Myr)/SFR($t$ $>$ 100 Myr) = 1 
   correspond to stellar populations forming continuously with a constant star 
   formation rate since 10 Gyr ago. Synthetic spectra labeled by $b$ $>$ 1
   correspond to stellar populations forming continuously between 0 and 10 Gyr
   with a star formation rate enhanced by a factor of $b$ during the last 100 Myr.
   These SEDs are superposed on the observed spectra uncorrected for extinction.
     }
    \label{fig:spfit_c2}
\end{figure}

\subsubsection{Continuous star formation with a young stellar population 
\label{young}}

To fit the observed SEDs of regions 1 -- 5  with only a young stellar 
population continuously formed over the last 100 Myr and derive the extinction,
we consider star formation occurring between $t_{\rm i}$ = 100 Myr and 
$t_{\rm f}$ = 8  Myr. This model predicts an $EW$(H$\delta$) = --7.2\AA, 
$EW$(H$\gamma$) = --5.9\AA, 
$EW$(H$\beta$) = 3.5\AA\ and $EW$(H$\alpha$) = 21.2\AA, 
close to the values observed in the LSB regions (Tables \ref{t:emhahb} and
\ref{t:abshdhg}). 
The results of our fitting are shown in Fig. \ref{fig:spfit_c1}.
We adjust the extinction coefficient $C$(H$\beta$) to achieve the
best agreement between the observed SED, after correction for interstellar 
extinction, and the theoretical SED. The observed extinction-corrected
SEDs are superimposed on the synthetic SEDs for regions 1 to 5 in Fig. 
\ref{fig:spfit_c1}a -- \ref{fig:spfit_c1}e. They are labeled by the derived 
values of $C$(H$\beta$). The synthetic SEDs reproduce the 
observed spectra quite well.
We therefore conclude that continuous star 
formation, occuring during the last 8 -- 100 Myr, is a reasonable
model of the star formation in the LSB regions.

We also considered the effect of metallicity on the age determination and
find that it is small. We fit the observed SEDs for regions 1 -- 5 with 
model SEDs calculated for $Z$ = $Z_\odot$/50 and assuming 
for each region the same $C$(H$\beta$) as in Fig. \ref{fig:spfit_c1}. 
The best fits are obtained with models characterised by continuous
star formation occuring during the last 8 -- 110 Myr, similar to the age
range of models with the heavy element mass fraction $Z$ = $Z_\odot$/20.

Note, that the extinction, derived from the best fit of the 
observed SEDs, is larger in the southeastern part of the galaxy (regions 
1 and 2) than in the northwestern one (regions 4 and 5).
The values for the extinction are similar to those 
in some well studied H {\sc ii} regions, e.g. in the LMC 
(Oey et al. \cite{Oey00}), open clusters in our Galaxy 
(Piatti, Bica \& Clari\'a \cite{Piatti2000}) and some BCDs 
(Guseva et al. \cite{Guseva2001}).

\subsubsection{Continuous star formation including an old stellar population 
\label{chb0}}

We consider next continuous star formation scenarios in which an old stellar 
population is present. For this, we adopt $C$(H$\beta$) = 0 and consider models
with constant and varying SFRs in the age interval between 0 and 10 Gyr. 
Specifically, for a varying SFR, we consider two periods of
star formation with constant but different SFRs, occurring in the age interval 
$\leq$ 100 Myr and $>$ 100 Myr.
To quantify the recent-to-past star formation rate ratio, we 
use the  parameter $b$ = 
SFR($t$ $\leq$ 100 Myr)/SFR($t$ $>$ 100 Myr) following Guseva et al. (2001).
First we consider models with
constant star formation during the whole 0 -- 10 Gyr range, 
i.e. models with $b$ = 1.
In Fig. \ref{fig:spfit_c2}a -- \ref{fig:spfit_c2}e we show such SEDs 
superimposed on the observed spectra of regions 1 -- 5 (labeled $b$ = 1). 
It is evident
that these models do not reproduce the observed SEDs. However, by
increasing the parameter $b$ we can fit the observed SEDs. 
These theoretical SEDs are labeled by $b$ $>$ 1 
in Fig. \ref{fig:spfit_c2}a -- \ref{fig:spfit_c2}e. The predicted
equivalent widths of the hydrogen emission and absorption lines are also 
in agreement with the observed ones in the case of $b$ $>$ 1. 
If non-negligible extinction is present
in regions 1 -- 5, then to fit the observed SEDs, the parameter $b$ 
should be further increased. In particular, if the extinction 
coefficient $C$(H$\beta$) = 0.33 as derived for region 2 from the 
H$\alpha$/H$\beta$ flux ratio is adopted, 
then the parameter $b$ should be as high as $\sim$ 100 to fit the observations.
However, if $C$(H$\beta$) = 0.33 is assumed for region 4, then this model cannot fit the
observations, even for a $b$ = 1000. This implies, that 
$C$(H$\beta$) is likely to be small in region 4 (Fig.~\ref{fig:spfit_c1}).
Slightly lower values (by $\sim$ 10\%) of the parameter $b$ are 
obtained if theoretical SEDs with $Z$ = $Z_\odot$/50 are used instead of the 
models with $Z$ = $Z_\odot$/20. 

In summary, analysis of the spectroscopic data for the 
LSB regions shows that the stellar population can be equally well reproduced by a 
model of continuous star formation  with constant SFR during the last 100 Myr, 
or by a model in which stars are continuously formed in the period 
0 -- 10 Gyr with varying SFRs. In the former case, a non-negligible extinction 
should be taken into account. In the latter case, a fit to the observed SED is 
only possible when $b$ $>$ 1 ($b$ = 6 -- 50, if $C$(H$\beta$) = 0 and 
$b$ $>$ 100, if $C$(H$\beta$) $>$ 0.2 -- 0.3), i.e. the star formation rate in the 
LSB component
has significantly increased over the last 100 Myr. 

\subsection{Age from the colour distribution \label{coldist}}

Photometric data allow us to check the results obtained in Sect. 5.1 through 5.3.
For this purpose we extract from the $V$ and $I$ images the same 
areas as those covered by the long slit spectroscopic observations and compare the
$(V-I)$ colours with predictions from our population synthesis modeling.

\begin{figure}[hbtp]
              \psfig{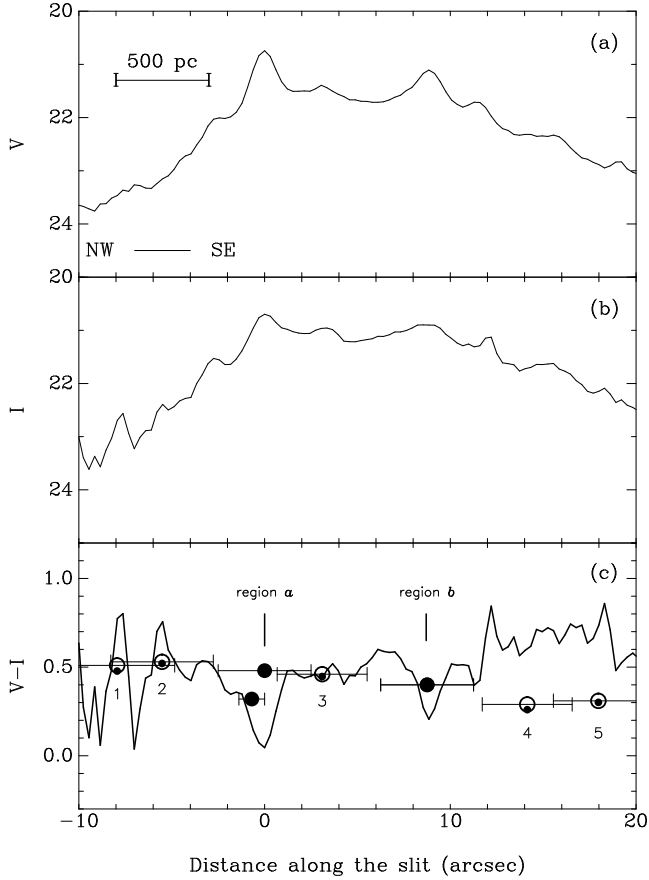}
    \caption{({\bf a}) and ({\bf b}) Surface brightness distributions along the
slit in $V$ and $I$. The origin is at the location of region {\it a}. 
({\bf c})~$(V-I)$ colour distribution along the slit. The two bright 
H {\sc ii} regions are marked as ``region {\it a}'' and ``region {\it b}''.
The LSB regions  
are marked as ``1'' to ``5''.
Small filled circles show the predicted colours of a
stellar population formed continuously with a constant star formation rate
between 8 Myr and 100 Myr ago and reddened with an extinction coefficient as 
derived for 5 regions  (see  Fig. \ref{fig:spfit_c1}).
Open circles indicate the predicted colours of stellar populations
formed continuously between 0 and 10 Gyr with the enhancement parameters
$b$ as derived for  these regions (see Fig. \ref{fig:spfit_c2}). 
Large filled circles show the colours for a stellar population formed in 
an instantaneous burst with ages $t$ = 8 and 10 Myr
for regions {\it a} and {\it b}, respectively. These ages are derived from the 
observed equivalent widths of the H$\alpha$ and H$\beta$ emission lines.
 }
    \label{fig:colsect1}
\end{figure}

The results of this comparison are shown 
in Fig.~\ref{fig:colsect1}. With solid lines we show in panels (a) and (b)
the observed $V$ and $I$ surface brightness distributions in mag 
arcsec$^{-2}$ along the slit while the corresponding $(V-I)$ colour 
distribution is shown in panel (c).
The predicted colours, obtained from convolved theoretical spectral energy
distributions with the appropriate filter bandpasses 
are shown in Fig.~\ref{fig:colsect1}c with different symbols.
The transmission curves for Johnson $V$  and Cousins $I$ bands are taken
from Bessell (\cite{B90}) and the zeropoints are from Bessell et al. (\cite{B98}). 

Since the contribution of the gaseous emission is 
very small in  regions 1 -- 5, we do not take it into account 
 and consider the colours of a stellar population 
formed continuously with a constant star formation rate between 
8 Myr and 100 Myr ago and reddened by interstellar 
extinction. In Fig.~\ref{fig:colsect1}c we show by small filled circles the
colours of five regions with extinction coefficients derived from the best
theoretical fit to the observed extinction-corrected spectra 
(see spectra in Fig. \ref{fig:spfit_c1}a -- \ref{fig:spfit_c1}e).
Open circles show the predicted colours of a stellar population continuously 
formed between 0 and 10 Gyr ago, assuming an enhanced star formation during the
last 100 Myr, as defined by the parameter $b$ (see upper spectra
in Fig. \ref{fig:spfit_c2}a -- \ref{fig:spfit_c2}e). 
In this case $C$(H$\beta$) = 0 is adopted.

The contribution of the ionized gas emission to the integrated light
of regions {\it a} and {\it b} is small but not negligible. We have, therefore,
included gaseous emission in our calculations.
We have constructed stellar population SEDs for a heavy element mass
fraction $Z$ = $Z_{\odot}$/20 and instantaneous burst ages of 8 and 10 Myr 
for regions {\it a} and {\it b} respectively as
derived from $EW$(H$\alpha$) (Fig.~\ref{fig:ewhbha}) and 
$EW$(H$\beta$).
The calculation of the gaseous SED
is made in the same way as in Guseva et al. (2001) and 
then added to the SED of the stellar component.
The derived colours are shown in Fig.~\ref{fig:colsect1}c by 
large filled circles.
The colour varies sharply near the peak of intensity in regions {\it a} and 
{\it b}. The agreement between the observed and calculated colours in the 
wide apertures is fair, whereas it is better in the small aperture of region 
{\it a}.

The agreement between the $(V-I)$ colours obtained from the photometric 
data and those derived from the spectral energy distributions in both 
scenarios is  quite good, 
given 
the large extraction apertures
for regions along the slit and a possible small positional shift between  
photometric and spectroscopic data.
The systematically redder colours observed at the relative position
of regions 4 and 5 are likely caused by 
the red background galaxies such as 
 {\it bg}  (Fig.~\ref{f1}), intersecting the slit in the 
northwestern part of the galaxy. 
This emphasizes the need for identification and rejection of background 
or foreground sources when colours are used to derive ages.
Figure 1 shows a sizable number of red background galaxies all over the 
field of SBS 1129+576 spanning a range of about 10 mag with respect to
their $I$ magnitudes. 
Therefore, the failure of any continuous star formation model 
to reproduce the observed red colours in regions 4 and 5 is not surprising.
Note that the average ($V$--$I$) colour in the outer part of SBS 1129+576 
does not exceed
0.6 mag, i.e. it is bluer by $\ga$ 0.2 mag than the observed colour 
in regions 4 and 5.

 \section{Conclusions \label{conc}}

We present a detailed spectroscopic and  photometric study
of the very metal-deficient dwarf irregular galaxy SBS~1129+576,
a likely young galaxy candidate. Broad-band $V$ and $I$ imaging 
and  spectra in the optical range have been obtained 
with the Kitt Peak 2.1m and 4m telescopes, respectively.
The main conclusions of this study can be summarized as follows:

\begin{enumerate}

\item SBS~1129+576 is a very low-metallicity nearby ($D$ = 26.3 Mpc) dwarf 
galaxy with a chain of H {\sc ii} regions along its elongated 
stellar low-surface brightness (LSB) component. 
The average $(V-I)$ colour of the outer part of the galaxy
with a surface brightness  $\mu$$_V$ in the range 23~--~26 mag arcsec$^{-2}$ 
is relatively blue $\sim$0.56$\pm$0.03 mag,
as compared to ($V-I$) $\sim$0.9--1.0 in the majority of dwarf irregular 
and blue compact dwarf  (BCD) galaxies.
The scale length $\alpha$ obtained from the surface brightness 
profiles in the $V$ and $I$ is $\sim$ 430 pc.

\item The oxygen abundance is found to be respectively
12 + log(O/H) = 7.36 $\pm$ 0.10 and 7.48 $\pm$ 0.12
in the two brightest H {\sc ii} regions (regions {\it a} and {\it b}).
Because of the low intensity of the [O {\sc iii}]$\lambda$4363
in region {\it a}, shock enhancement of this line may be important, 
which may result in a slight underestimate of the oxygen abundance. 
The neon-to-oxygen abundance ratio log~Ne/O = --0.76  in region {\it a}  
is in good agreement with the mean ratio 
derived from the previous studies of low-metallicity galaxies (e.g., 
Izotov \& Thuan \cite{IT99}).
The nitrogen-to-oxygen abundance ratio log N/O = --1.60 lies in the narrow 
range of the N/O ratios obtained by Thuan et al. (\cite{til95}) and Izotov \& 
Thuan (\cite{IT99}) for the most metal-deficient BCDs. 

\item The He mass fractions $Y$ = 0.220 $\pm$ 0.027 and 0.202 $\pm$ 0.042
derived, respectively, in regions {\it a} and {\it b} 
are significantly lower than the value of the primordial He 
mass fraction $Y_{\rm p}$ = 0.244 -- 0.245 derived previously by Izotov \& Thuan 
(\cite{IT98a}) and Izotov et al. (\cite{ICFGGT99}). This difference
is likely due to significant underlying He {\sc i} stellar 
absorption in SBS 1129+576. Hence, despite its low metallicity, 
this galaxy is not a good candidate for primordial helium determination. 

\item Hydrogen H$\alpha$ and H$\beta$ Balmer lines 
are seen in emission in the LSB component, while higher order 
hydrogen lines are in absorption.
Two extinction-insensitive
methods, based on the 
temporal evolution of the H$\alpha$ and H$\beta$ emission line
and the H$\gamma$ and H$\delta$ absorption line equivalent widths are used
for age determination.
A third method (not extinction-independent) is based on the age 
dependence of the spectral energy distribution. 
Several star formation histories have been considered. 
We find that models of continuous star formation with a constant star
formation rate starting 10 Gyr ago are excluded. However, models 
starting 10 Gyr ago and continuing to the present with a varying star formation 
rate are able to account for the observed properties of the
LSB regions.  
Models with star formation rates 
enhanced by 
6 -- 50 times during the last 100 Myr 
can reproduce the observed equivalent widths of the emission and absorption
hydrogen lines and SEDs if zero extinction is assumed. 
If some extinction is present in the  LSB component then 
the star formation rate during the last 100 Myr
should be enhanced by a factor of more than 100 times, or
alternatively
the observed spectroscopic and photometric characteristics 
of the LSB component can be 
reproduced 
by models in which
stars were continuously formed during the last 100 Myr only.

\item The observed $(V-I)$ colours in the LSB component of the galaxy 
are consistent with colours of synthetic SEDs for all above mentioned histories of 
star formation. Hence, we conclude, that there is no compelling evidence for 
either a young or an old age of SBS 1129+576.

\end{enumerate}

\begin{acknowledgements}
N.G.G. thanks the support of DFG grant 436 UKR 17/2/02 and Y.I.I. 
is supported by a Gauss professorship
of the G\"{o}ttingen Academy of Sciences.
They also thank Swiss SCOPE 7UKPJ62178
grant and the hospitality at G\"ottingen Observatory. Y.I.I. and 
T.X.T. have been partially supported by NSF grant AST-02-05785.
Research by P.P. and K.J.F. has been supported by the
Deutsches Zentrum f\"{u}r Luft-- und Raumfahrt e.V. (DLR) under
grant 50\ OR\  9907\ 7.  K.G.N. thanks the support from the Deutsche
Forschungsgemeinschaft (DFG) grants FR 325/50-1 and FR 325/50-2.
This research has made use of the NASA/IPAC
Extragalactic Database (NED) which is operated by the Jet Propulsion
Laboratory, California Institute of Technology, under contract
with the National Aeronautics and Space Administration.
\end{acknowledgements}


\begin{thebibliography}{}


\bibitem[1984]{Aller84} Aller, L. H. 1984, Physics of Thermal Gaseous Nebulae,
   Dordrecht: Reidel

\bibitem[1989]{Anders89} Anders, E., \& Grevesse, N.
  1989, Geochim.Cosmochim.Acta, 53, 197

\bibitem[1994]{Bertelli94} Bertelli, G., Bressan, A., Chiosi, C., Fagotto, F., 
\& Nasi, E. 1994, A\&AS, 106, 275

\bibitem[1990]{B90} Bessell, M. S. 1990, PASP, 102, 1181

\bibitem[1998]{B98} Bessell, M. S., Castelli, F., \& Plez, B. 1998, A\&A, 
333, 231

\bibitem[1986]{Bica86} Bica, E., \& Alloin, D. 1986, A\&A, 162, 21


\bibitem[2000]{b2000} Bicay, M. D., Stepanian, J. A., Chavushyan, V. H., et al.
2000, A\&AS, 147, 169

\bibitem[1991]{binggeli91} Binggeli, B., \& Cameron, L. M. 1991, A\&A, 252, 27


\bibitem[1971]{Brocklehurst71} Brocklehurst, M. 1971, MNRAS, 153, 471

\bibitem[1993]{Caon93} Caon, N., Capaccioli, M., $\&$ D'Onofrio, M. 1993, MNRAS, 265, 1013

\bibitem[1994]{cellone94} Cellone, S. A., Forte, J. C., \& Geisler,
D. 1994, ApJS, 93, 397

\bibitem[2000]{Cervino00} Cervi\~no, M., Luridiana, V., \& Castander, F. J. 
2000, A\&A, 360, 5

\bibitem[1996]{F96} Ferland, G. J. 1996, HAZY, A Brief Introduction to CLOUDY 
90 (University of Kentucky, Department of Physics and Astronomy, Internal 
Report)

\bibitem[1998]{F98} Ferland, G. J., Korista, K. T., Verner, D. A., et al. 
1998, PASP, 110, 761

\bibitem[1997]{F97} Fioc, M., \& Rocca-Volmerange, B. 1997, A\&A, 326, 950

\bibitem[2001]{Fricke00} Fricke, K. J., Izotov, Y. I., Papaderos, P., 
Guseva, N. G., \& Thuan, T. X. 2001, AJ, 121, 169

\bibitem[1999]{GonLeith99b} Gonz\'alez Delgado, R. M., Leitherer, C., \&
 Heckman, T. M. 1999, ApJS, 125, 489

\bibitem[2001]{Guseva2001} Guseva, N. G., Izotov, Y. I., Papaderos, P., et al.
2001, A\&A, 378, 756

\bibitem[1998]{IT98a} Izotov, Y. I., \& Thuan, T. X. 1998, ApJ, 500, 188 

\bibitem[1999]{IT99} Izotov, Y. I., \& Thuan, T. X. 1999, ApJ, 511, 639

\bibitem[1994]{ITL94} Izotov, Y. I., Thuan, T. X., \& Lipovetsky, V. A. 1994, 
 ApJ, 435, 647  

\bibitem[1997]{Izotov97} Izotov, Y. I., Thuan, T. X., \& Lipovetsky, V. A. 
1997, ApJS, 108, 1

\bibitem[1999]{ICFGGT99} Izotov, Y. I., Chaffee, F. H., Foltz, C. B., et al.
1999, ApJ, 527, 757


\bibitem[1992]{Landolt92} Landolt, A. U. 1992, AJ, 104, 340

\bibitem[1998]{Lejeune98} Lejeune, T., Cuisinier, F., \& Buser, R.
   1998, A\&AS, 130, 65


\bibitem[1988]{l88} Lipovetsky, V. A., Stepanian, J. A., Erastova, L. K.,
\& Shapovalova, A. I. 1988, Afz, 29, 548


\bibitem[1999]{m99} Makarova, L. 1999, A\&A, 139, 491

\bibitem[1983]{ms83} Markarian, B. E., \& Stepanian, J. E. 1983, Afz, 19, 639


\bibitem[1983]{Mendoza83} Mendoza, C. 1983, in IAU Symp. 103, Planetary 
Nebulae, ed. Flower D. R. (Dordrecht:Reidel), p. 143


\bibitem[2000]{Noeske00} Noeske, K. G., Guseva, N. G., Fricke, K. J., et al.
2000, A\&A, 361, 33

\bibitem[2000]{Oey00} Oey, M. S., Dopita, M. A., Shields, J. C., \& 
Smith, R. C. 2000, ApJS, 128, 511

\bibitem[1996a]{Papa96a} Papaderos, P., Loose, H.-H., Thuan, T. X., \&
  Fricke, K. J. 1996a, A\&AS, 120, 207

\bibitem[1996b]{Papa96b} Papaderos, P., Loose, H.-H., Fricke, K. J., \&
Thuan, T. X. 1996b, A\&A, 314, 59

\bibitem[2002]{papaderos02}Papaderos, P., Izotov, Y. I., Thuan, T. X., et al.
2002, A\&A, 393, 461

\bibitem[1996]{patterson96} Patterson, R. J., \& Thuan, T. X. 1996, ApJS, 107, 
103

\bibitem[2000]{Piatti2000} Piatti, A. E., Bica, E., \& Clari\'a, J. J.
  2000, A\&A, 362, 959

\bibitem[1994]{RB94} R\"onnback, J., \& Bergvall, N. 1994, A\&AS, 108, 193


\bibitem[1999]{SN99} Salzer, J. J., \& Norton, S. A. 1999, in Low Surface 
Brightness Universe, ASP Conference Series 170, Eds J. I. Davies, C. Impey, 
and S. Phillipps, p.253

\bibitem[1968]{sersic68}S\'ersic, J.--L. 1968, Atlas de Galaxias
Australes, Observatorio Astronomico de Cordoba

\bibitem[1990]{Stasinska90} Stasi\'nska, G. 1990, A\&AS, 83, 501

\bibitem[2003]{StasinskaIz02} Stasi\'nska, G., \& Izotov Y. I. 2003, A\&A,
397, 71
 
\bibitem[1997]{telles97} Telles, E., \& Terlevich, R. 1997, MNRAS, 286, 183

\bibitem[1995]{til95} Thuan, T. X., Izotov, Y. I., \& Lipovetsky, V. A. 1995, 
  ApJ, 445, 108


\bibitem[1999]{tlmp99} Thuan, T. X., Lipovetsky, V. A., Martin, J.-M., \&
Pustilnik, S. A. 1999, A\&AS, 139, 1

\bibitem[1996]{Vacca96} Vacca, W. D., Garmany, C. D., \& Shull, J. M. 1996, 
ApJ, 460, 914

\bibitem[2000]{v2000} van Zee, L. 2000, AJ, 119, 2757

\bibitem[1996]{vennik96} Vennik, J., Hopp, U., Kovachev, B., et al. 1996,
A\&AS, 117, 216

\bibitem[2000]{vennik00} Vennik, J., Hopp, U., \& Popescu, C. C. 2000, 
A\&AS, 142, 399

\end{thebibliography}
\end{document}